\documentclass{aa}

\newcommand{\teff}{\mbox{$T_{\mathrm{eff}}$}}
\newcommand{\teffp}{\mbox{$T_{\mathrm{eff,1}}$}}
\newcommand{\teffs}{\mbox{$T_{\mathrm{eff,2}}$}}

\newcommand{\loggp}{\ensuremath{\log g_{\mathrm 1}}}
\newcommand{\loggs}{\ensuremath{\log g_{\mathrm 2}}}
\newcommand{\vsini}{\mbox{$\upsilon \sin i$}}

\def\kms{$\mathrm{km\,s}^{-1}$}

\newcommand{\porb}{\mbox{\ensuremath{P_{\mathrm{orb}}}}}

\newcommand{\rsun}{R\ensuremath{_\odot}}
\newcommand{\msun}{M\ensuremath{_\odot}}

\newcommand{\degree}{\mbox{\ensuremath{^\circ}}}   

\newcommand{\mprim}{\ensuremath{M_{\mathrm {1}}}}
\newcommand{\msec}{\ensuremath{M_{\mathrm {2}}}}
\newcommand{\rprim}{\ensuremath{R_{\mathrm {1}}}}
\newcommand{\rsec}{\ensuremath{R_{\mathrm {2}}}}

\usepackage{graphicx,natbib,hyperref}
\usepackage{tikz,xcolor}

\hypersetup{
    colorlinks = true,
   linkcolor = {blue}, 
    citecolor = {blue},
    urlcolor = {blue}
}

\usepackage{txfonts}

\usepackage[normalem]{ulem}

\begin{document}

   \title{Discovery of the pre-main-sequence eclipsing binary \object{MML 48}}

   \author{Y.\ G\'omez Maqueo Chew\inst{\ref{ia}}   
       \and L.\ Hebb\inst{\ref{hws}}
          \and  H.C.\ Stempels\inst{\ref{upp}}
          \and F.\ M.\ Walter \inst{\ref{stony}}
          \and D.\ J.\ James\inst{\ref{d1},\ref{d2}}
          \and G.\ A.\ Feiden\inst{\ref{ung}}
         \and R. Petrucci\inst{\ref{unc},\ref{conicet}}
         \and T.\ Lister\inst{\ref{lco}}
         \and I.\ Baraffe\inst{\ref{exeter},\ref{lyon}}
         \and M.\ Brodheim\inst{\ref{hws},\ref{keck}} 
         \and F. Faedi\inst{\ref{spacepark}} 
         \and D.R.\ Anderson\inst{\ref{keele}} 
         \and R.A.\ Street\inst{\ref{lco}}
          \and C. Hellier\inst{\ref{keele}}
          \and K.G. Stassun\inst{\ref{vandy}}
          }

   \institute{Universidad Nacional Aut\'onoma de M\'exico,  Instituto de Astronom\'ia, Ciudad Universitaria, 04510 Ciudad de  M\'exico, M\'exico  \label{ia}
                \email{ygmc@astro.unam.mx}
                \and Department of Physics, Hobart and William Smith Colleges, Geneva, New York, 14456, USA\label{hws} 
                \and Department of Physics \& Astronomy, Uppsala University, Box 516, SE-75120 Uppsala, Sweden\label{upp}
                \and Department of Physics and Astronomy, Stony Brook University, Stony Brook, New York 11794, USA\label{stony} 
                \and ASTRAVEO LLC, PO Box 1668, MA 01931, USA \label{d1}
                \and Applied Materials Inc., 35 Dory Road, Gloucester, MA 01930, USA \label{d2}
                \and Department of Physics \& Astronomy, University of North Georgia, 82 College Circle, Dahlonega, GA 30597, USA \label{ung}
                \and Universidad Nacional de Córdoba, Observatorio Astronómico de Córdoba, Laprida 854, X5000BGR Córdoba, Argentina \label{unc}
                \and Consejo Nacional de Investigaciones Científicas y Técnicas (CONICET), Godoy Cruz 2290, CABA, CPC 1425FQB, Argentina \label{conicet} 
                \and Las Cumbres Observatory Global Telescope Network, 6740 Cortona Dr. Suite 102, Goleta, CA 93117, USA  \label{lco}
                \and Physics and Astronomy, University of Exeter, Exeter EX4 4QL, UK  \label{exeter}
                \and  \'Ecole Normale Sup\'erieure, Lyon, CRAL (UMR CNRS 5574), Universit\'e de Lyon, France \label{lyon}
                \and W. M. Keck Observatory, Kamuela, HI 96743, USA \label{keck}
             \and  Institute for Space, Space Park Leicester, 92 Corporation Rd, Leicester LE4 5SP, UK \label{spacepark}
             \and Astrophysics Group, Keele University, Staffordshire, ST5 5BG, UK \label{keele}
             \and Department of Physics and Astronomy, Vanderbilt University, Nashville, TN 37235, USA \label{vandy} }

   \date{Received May 21, 2025; accepted July 17, 2025}

  \abstract 
  {}
   {We present the discovery of the eclipsing binary \object{MML 48}, which is a member of Upper Centaurus Lupus, has an associated age of 16 Myr, and is composed of two young, low-mass stars.}
   {We used space- and ground-based observations to characterize the system with both time-series photometry and spectroscopy. Given the extreme mass ratio between the stars, $q_{\rm EB}$ = 0.209  $\pm$  0.014, we modeled a single-lined spectroscopic and eclipsing binary system. }
   {The orbital period, 2.0171068 $\pm$ 0.0000004 d, is measured from the highest precision light curves. 
   We derive a primary mass of 1.2 $\pm$ 0.07 \msun\ using stellar models, and with radial velocities we measured a secondary mass of 0.2509 $\pm$ 0.0078 \msun. The radii are large, as expected for pre-main-sequence stars, and are measured as 1.574 $\pm$ 0.026 $\pm$ 0.050 \rsun\ and 0.587 $\pm$ 0.0095  $\pm$ 0.050 \rsun, for the primary and secondary stars, respectively. 
   }
   {MML~48 joins the short list of known low-mass, pre-main-sequence {  eclipsing binaries (EBs)},
being one of only five systems with intermediate ages (15-25 Myr), and the system with the most extreme mass ratio. The primary star is currently at the ``fusion bump", undergoing an over-production of energy in the core due to the build-up of  $^3$He before reaching its equilibrium abundance set by the {  proton-proton (p-p)} I chain. 
   {  MML~48~A is the first young star in an eclipsing system that has been found during its fusion bump.}
   MML~48 is thus an important benchmark for low-mass stellar evolution at a time when the stars are rapidly changing, which allows for a tight constraint on the corresponding isochrone given the uneven mass ratio. }


   \maketitle

\section{Introduction}

Mass and radius are the most fundamental stellar properties, and knowledge of these characteristics, for stars with a range of masses and ages, is key to understanding stellar evolution. Detached eclipsing binary stars (EBs), which are also double-lined spectroscopic binaries, provide accurate direct measurements (3\% precision) of fundamental stellar properties, including mass, radius, and temperature ratio \citep{Andersen1991,Torres2010, Kallrath2009}. Thus, EBs are essential for testing and calibrating theoretical stellar-evolution models, which describe our physical understanding of how stars form and evolve and are the primary means for deriving masses and ages of single stars. For this reason, theoretical models are used extensively in other areas of research to derive the stellar initial-mass function, to obtain star formation histories of local and distant stellar populations, and to determine the masses of exoplanets and their host stars.

Young EBs are particularly important because they constrain pre-main-sequence stellar evolution models \citep[e.g.,][]{Baraffe2015,Dotter2008} in the regime when the temperatures, luminosities, and radii of stars are changing rapidly as they settle onto the main sequence.
The number of young ($< 50$~Myr), low-mass ($<$ 1.4 \msun) EBs has more than doubled to a total of 20 systems in the last five~years 
\citep[see][]{GomezMaqueoChew2019}. Since 2019, the newly discovered and published double-lined, low-mass, young EBs are 
\object{Mon-735} \citep{Gillen2020}, 
\object{THOR-42} \citep{Murphy2020},
\object{USco 48} \citet{David2019}, 
\object{TOI-450} \citep{Tofflemire2022}, 
\object{2M1222-57} \citep{Stassun2022}, 
\object{2M05-06,} and \object{2M05-00} \citep{Kounkel2024}. 

Despite the growing number of young EBs that are known, it is important to note that these systems are {  found in five} star-forming regions: the Orion complex \citep{Covino2004,Stassun2004, Irwin2007, Stempels2008, GomezMaqueoChew2009, GomezMaqueoChew2012, Murphy2020, Kounkel2024}, the Scorpius-Centaurus {  (Sco-Cen)} complex \citep{Alonso2015, David2016, GomezMaqueoChew2019, David2019, Stassun2022}, NGC~2264 \citep{Gillen2014, Gillen2020}, Perseus \citep{Lacy2016}, and Columba \citep{Tofflemire2022}. 

In this paper, we present the discovery and preliminary analysis of the pre-main-sequence eclipsing binary MML~48, which is a member of Upper Centaurus Lupus {  (UCL)} in the Scorpius-Centaurus association \citep{Mamajek2002, Wright2018}, which has an age of 16~$\pm$~2~Myr \citep{Pecaut2012}. It was first identified to be eclipsing in the WASP light curves \citep{Pollacco2006}. Follow-up radial velocity {  (RV)} and photometric observations were acquired and are presented in Sect.~\ref{sec:observations}. Our data and analyses, presented in Sect.~\ref{sec:analysis}, show a single-lined, eclipsing binary composed of two young stars.  We discuss MML~48 in the context of the other known young EBs and compare it to stellar models in Sect.~\ref{sec:discussion}. 

More data are required to allow for the fundamental properties of the young MML~48 stars to be measured directly to robustly test stellar models and to study their magnetic fields, but these observations are challenging given the small mass ratio between the eclipsing components. And as such, these are beyond the scope of this discovery paper.

\section{Observations}
\label{sec:observations}

\subsection{Photometric data}

All light curves of MML~48 presented in this section are given in Table~\ref{tab:lcs}.

\begin{table}
\caption{Light-curve data for MML~48.}\label{tab:lcs}
\begin{tabular}{c c c c c c c}
\hline\hline\\
 BJD$_{\rm TDB}$      & $\Delta {\rm mag}$      & $\sigma_m$      & Telescope & Filter  \\
 $- 2\,450\,000$ & (kms$^{-1}$) & (kms$^{-1}$) & \phantom{telescope} &\phantom{Filter} \\
\hline\\
6476.921406     &        0.00036        & 0.00128 & FTS & z$_s$\\
6476.922484     &        0.00050        & 0.00129 & FTS & z$_s$\\
... & ... & ... & ... \\
\hline\\
\end{tabular} \\
{\footnotesize The full table is available at the CDS.}
\end{table}

\subsubsection{WASP photometry} 

The EB MML~48 was observed in the{  field of view} of the Wide Angle Search for Planets (WASP) transiting planet survey \citep{Pollacco2006} from 2006-2014.  WASP is a wide-field photometric variability survey designed to detect transiting gas-giant planets around bright main-sequence stars ($V\sim 9-13$).  The standard WASP data reduction and photometry pipeline \citep{hunter} was applied to all fields containing MML~48.  The resulting single-band (wide V+R filter) light curves contain a total of 105,609 individual photometric measurements in seven different seasons between 2006 and 2014.   
However, no observations of MML~48 were made in 2009-2010.   
The typical photometric precision of the 2006–2012 data is $\sim 7$~mmag with a cadence of $\sim 8$~minutes.  After upgrading the camera lenses in July 2012 and applying a new observing strategy to reduce correlated noise, the photometric precision of the 2013-2014 data dropped to $\sim 30$~mmag, but the cadence increased to $< 2$~minutes.  

As part of the standard WASP analysis pipeline, the box least-squares algorithm \citep{Kovacs2002, hunter} was applied to the light curves of MML~48, and the system was flagged as a likely EB with a period of $\sim2$~days.  The pipeline applies a photometric aperture of $48^{\prime\prime}$ which contains one other star of comparable brightness (2MASS~14413595-4700280), which is approximately $15^{\prime\prime}$ to the west of MML~48 and 2.56~magnitudes fainter in the Gaia G-band \citep{Gaia2016,GaiaDR32023}.  Subsequent follow-up photometry excluding this target shows that the eclipsing object is indeed MML~48, but the contamination means that the WASP light curves are not suitable for the EB analysis used to derive the fundamental properties of the component stars.  Finally, despite the close visual proximity of this fainter star, its very different Gaia{  Data Release 3} parallax (${\rm plx} = 0.1150 \pm 0.0208~mas$) compared to MML~48 (${\rm plx} = 8.7666 \pm 0.0190~mas$) suggests it is a line-of-sight background star that is unassociated with MML~48 or the young Sco-Cen star forming region.  

\begin{figure}
\centering
\includegraphics[width=\columnwidth]{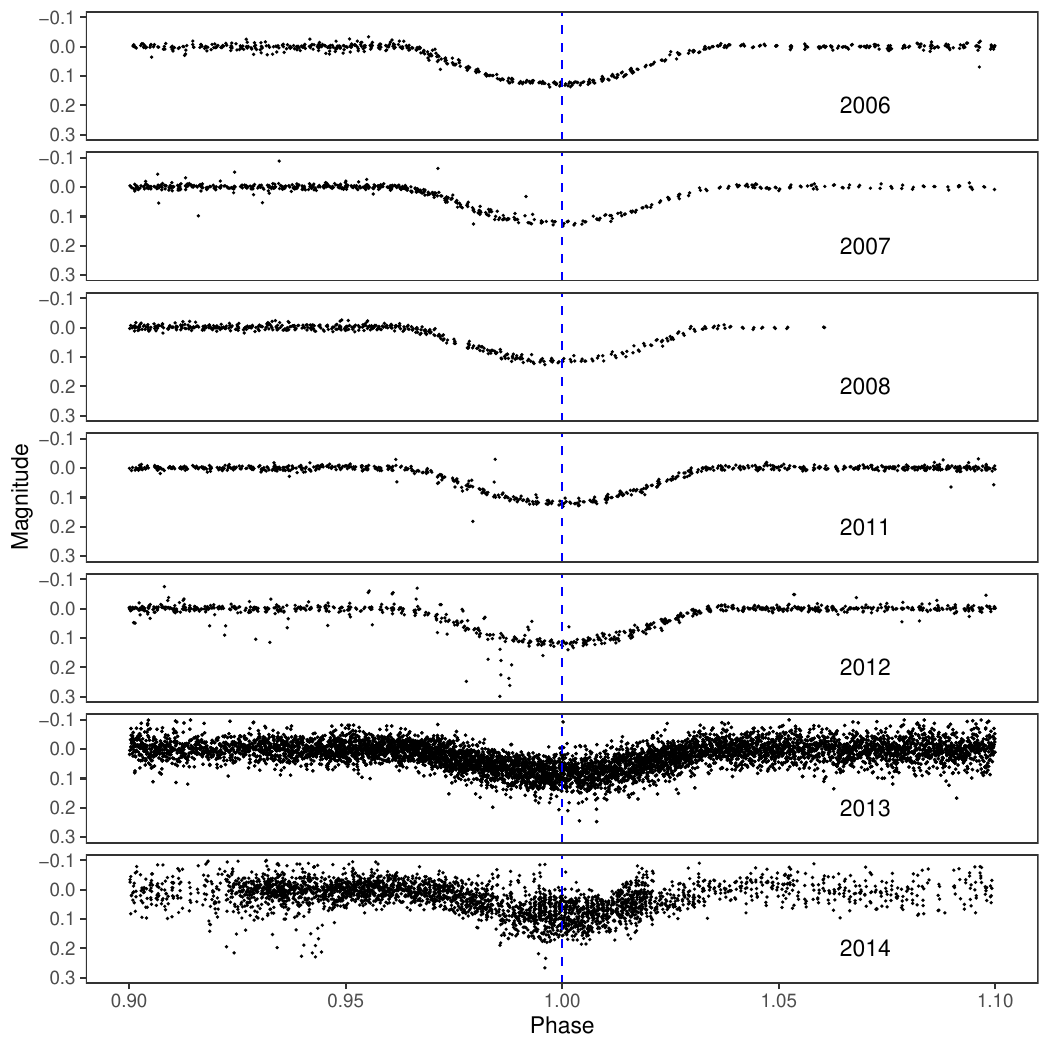}
\caption{\label{fig:swasp} WASP time-series photometry of primary eclipses of  MML 48 obtained in 2006--2014.  
The secondary eclipse is too shallow to see in this photometry, indicating the secondary is a very low-mass companion. These data were not used to derive the EB parameters since the light curves are contaminated by a nearby background star {  15$^{\prime\prime}$} to the west (2MASS~14413595-4700280). }
\end{figure}

Any night showing a full eclipse with out-of-eclipse photometry on both sides of the eclipse was extracted from the full WASP light curve and used to measure the epoch of minimum light of the eclipsing stars for each year of data between 2006--2014 as described in Sect.~\ref{sec:timings}.  If no full eclipses were captured in a season, the best partial eclipses were used for this same purpose.   Starspot modulation is present in the WASP light curves, and the phase of the variation persists over multiple seasons suggesting the spots are large and long-lived.  This is typical for young, rapidly rotating stars \citep[e.g.,][]{gj1243}.
However, the rotational modulation can affect the derivation of eclipse times if not modeled correctly or removed.  Therefore, each night of data was rectified by fitting a low-order 
polynomial to the out-of-eclipse data and subtracting the model values from the observed magnitudes at all times. Individual out-of-eclipse data points were rejected at this stage if they deviated by more than $5\sigma$ from the polynomial baseline. The resulting phase-folded primary eclipses derived from the rectified light curves are shown in Figure~\ref{fig:swasp} for each year of WASP data.  We omitted the July 2012 data since no eclipses occurred at night during this time period.   Finally, there is a lack of visible secondary eclipses in the white-light WASP filter, suggesting that the binary consists of two highly unequal-mass stars; subsequent follow-up \'echelle spectroscopy confirmed this (Sect.~\ref{sec:starchar}).

\subsubsection{LCOGT photometry}\label{sec:fts} 
\begin{figure*}[ht]
\centering
\includegraphics[width=0.475\hsize]{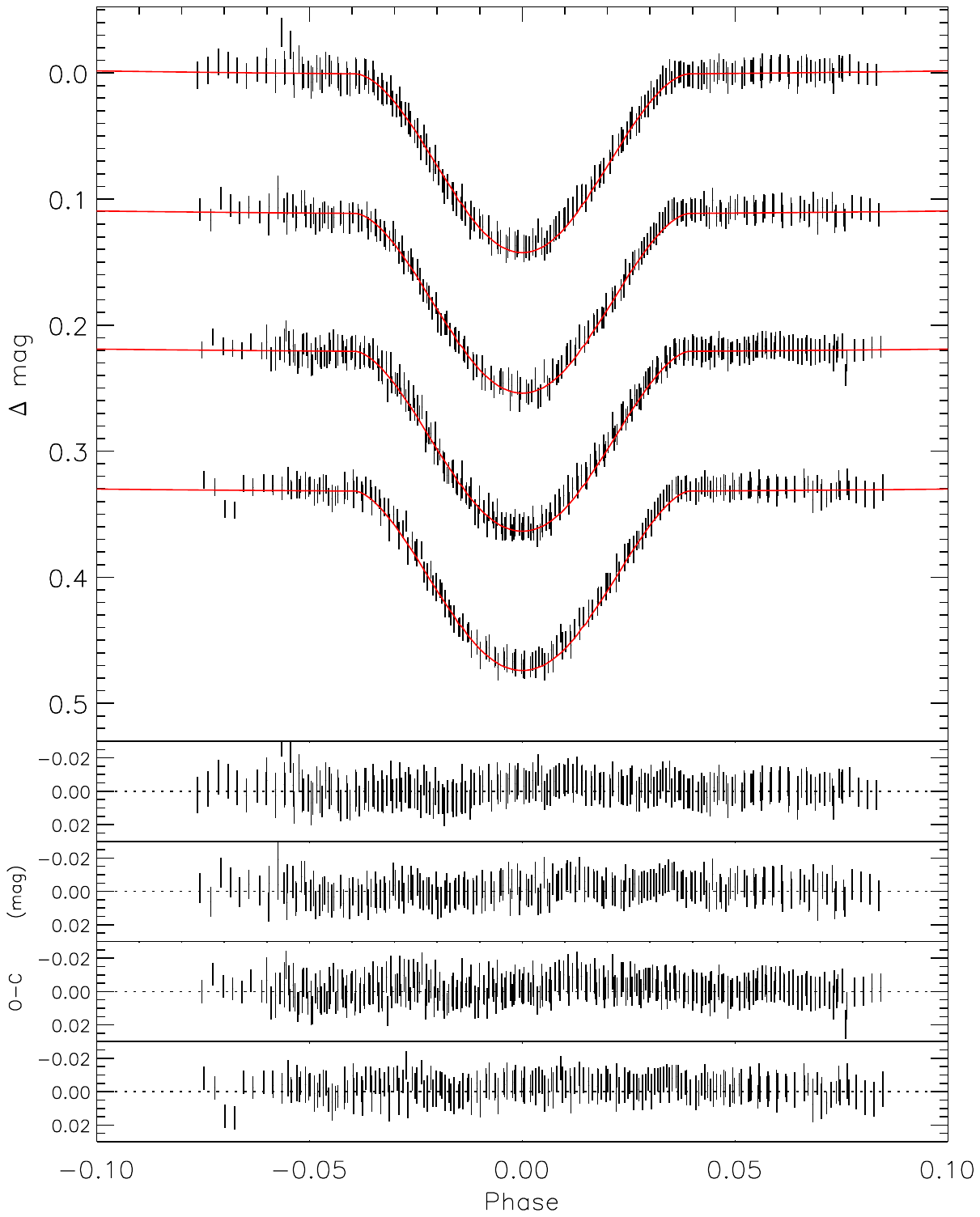}
\includegraphics[width=0.475\hsize]{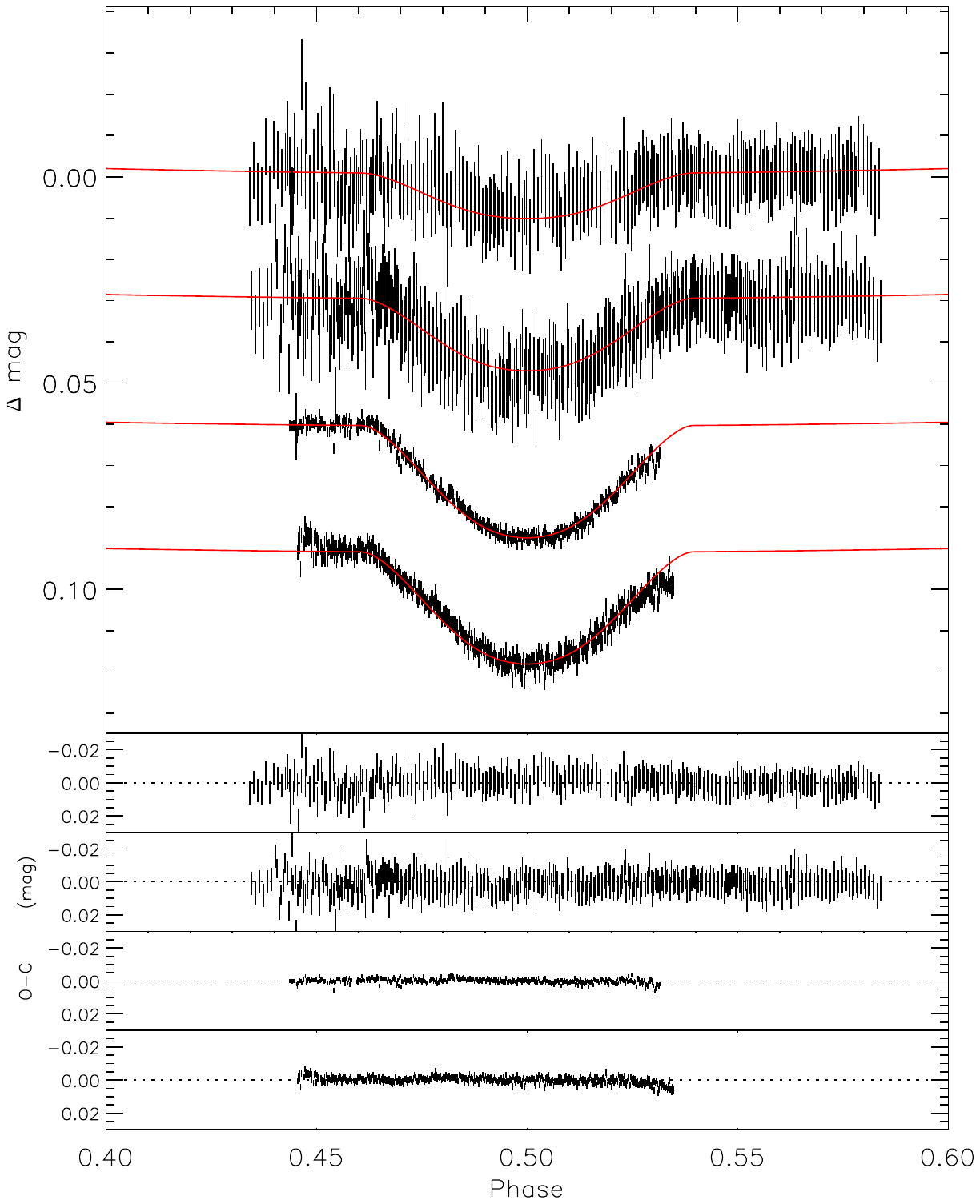}
\caption{{  P}rimary- and secondary-eclipse light curves of MML~48. Left: Primary-eclipse  light curves acquired at CTIO/SMARTS-0.9m telescope in BVRI bands (from top to bottom), each shifted vertically for clarity. Overplotted is the {\sc Phoebe} best fit model in the corresponding filter, shown by the continuous red line. 
The photometric uncertainties are represented by the length of the vertical lines. 
The residuals to the fit are shown in the bottom panels for each band. 
Right: Secondary-eclipse light curves acquired in the RI bands at the CTIO/SMARTS-0.9m (top two), and the FTS z'-band light curves (bottom two). The light curves are plotted on a different scale between the right and left panels.}
\label{fig:lcsfit}
\end{figure*}

Two secondary eclipses of MML~48 were observed on 3 July, 2013 and 5 July, 2013 in order to constrain the eccentricity of the orbit and the size and brightness of the secondary star through the timing and depth of the eclipse.  These data were obtained with the Spectral Camera on the 2-m Faulkes Telescope South (FTS) through the Las Cumbres Observatory Global Telescope Network (LCOGT). 
We observed the target with ten-second exposure times repeatedly for approximately 4.3 hours in the PanSTARRS z$_s$ filter.  We employed the $2\times2$~binning mode for faster readout time and defocused the camera by 0.55~mm to avoid saturation.
The data were processed in the standard way with the LCOGT imaging data pipeline (BANZAI),
\footnote{\url{https://lco.global/observatory/data/BANZAIpipeline/}} which includes bad pixel masking, bias and dark-frame subtraction, and flat-field division of each individual science frame with the best available calibration images.  Source detection and aperture photometry were performed on all processed science images using the Cambridge Astronomical Survey Unit{  catalog's} extraction software \citep{IrwinLewis2001}. 
Adopting conservative parameters
to define the detection threshold, the target star and dozens of fainter stars in the field were detected in each image.  Aperture photometry was performed on all detected stars using a five-pixel-radius circular aperture, which was selected to match the defocusing of the telescope.
The same aperture was used on both nights of data.  Thirteen bright,{  nonvariable} reference stars were selected from the many detected stars and used to perform differential photometry on the target star.  In each image, the flux from all reference stars was summed into a single, super comparison star that was divided by the aperture flux from MML~48 and converted to a differential magnitude.  The resulting phase-folded differential photometry light curves obtained from both nights of data are shown in Figure~\ref{fig:lcsfit}.

\subsubsection{CTIO/SMARTS 0.9 m{  p}hotometry}\label{sec:ctiophot} 

To obtain simultaneous multiband photometry suitable for detailed EB analysis, MML~48 was observed for six consecutive nights from 16-22 March, 2019 with the optical CCD camera on the Cerro Tololo Inter-American Observatory/Small and Moderate Aperture Research Telescope System (CTIO/SMARTS) 0.9 m telescope. 
The CCD camera has been consistently in place on this telescope since at least 1999 to enable long-term photometric and astrometric studies of nearby stars for the REsearch Consortium On Nearby Stars \citep[RECONS;][]{Henry2018}.\footnote{The RECONS Project at: \url{http://www.recons.org/}\label{recons}}  The camera is equipped with a $2048 \times 2048$ detector from Scientific Imaging Technologies, Inc. (SITe), which is split into four $1024 \times 1024$ quadrants, each one read by a single amplifier.   The $24\mu$ square pixels result in a plate scale of 401 mas/pixel and a field of view of $13.6^\prime$ square.  The device  has a typical full-well capacity of 200,000 e$^{-}$ --which limits the exposure time on bright stars-- and a read noise of 5-7 e$^{-}$.

The near two-day period of MML~48 allowed us to continuously  observe the primary eclipse {  throughout the night} on nights two, four, and six of our observing run %
in the Johnson-Cousins B, V, R, and I bands, while the secondary eclipse was observed on alternate nights in R and I only.  Exposure times were between 5 and 40~seconds depending on the filter, with the B band having the longest exposure times.  We only used one quadrant with a single amplifier; thus, the readout time was 0.8 minutes.  The raw images were processed with {\sc AstroImageJ}\footnote{https://www.astro.louisville.edu/software/astroimagej/} \citep{Collins2017} using nightly biases and dome flats with the appropriate filter. 
Aperture photometry was performed on MML~48 and six bright,{  nonvariable} comparison stars using a ten-pixel radius aperture.   The flux from all comparison stars was summed into a single, super comparison star that was divided by the aperture flux from MML~48 and converted to a differential magnitude.  The resulting phase-folded differential-photometry light curves obtained in all filters are shown in Figure~\ref{fig:lcsfit}.

\subsubsection{CASLEO photometry}\label{sec:casleo} 

The EB MML~48 was observed in the Johnson VRI bands during four consecutive nights (27-30 June, 2018) with the 2.15-m ``Jorge Sahade" telescope at Complejo Astron\'omico El Leoncito (CASLEO) in Argentina (PI: Petrucci; DDT). We used the Roper Scientific camera and a focal reducer that provide a 9\arcmin-diameter circular{  field of view} and a plate scale of 0.45\arcsec/pixel.
Exposure times (5--17~s) were chosen to maximize the flux in the target and nearby comparison stars while keeping the peak count value of the brightest star below 40\,000 counts. 
The telescope was kept focused during the observations to minimize contamination from the nearby star 2MASS~14413595-4700280. 

The images were processed with {\sc IRAF,}\footnote{The IRAF is distributed by the National Optical Astronomy Observatories, which are operated by the Association of Universities for Research in Astronomy, Inc., under cooperative agreement with the National Science Foundation.} correcting for bias and flat field. We used {\sc FOTOMCAp}  \citep{Petrucci2016} to perform aperture photometry of the stars in the processed images using the aperture-correction method \citep{Howell1989,Stetson1990}. 
The final differential light curve per filter was obtained by adding the flux from the reference stars with the lowest dispersion into a single, super comparison star that was divided by the aperture flux from MML~48. The MML~48 CASLEO light curves were not used in the EB analysis because the observations did not have enough baseline before the eclipses to properly correct for systematics. Additionally, the variable seeing during our observations caused dilution from 2MASS~14413595-4700280 in some exposures. In the most egregious cases, we discarded those images from the light curves. We show the CASLEO VRI light curves of the primary eclipses in the right panel of Fig.~\ref{fig:lcstess} for reference, with the best model overplotted, and we provide them as computer-readable tables to the community.

\subsubsection{TESS light curves}\label{sec:tess} %

\begin{figure}
\centering
\includegraphics[width=\columnwidth,trim=0.22cm 0.25cm 1.57cm 1.52cm, clip]{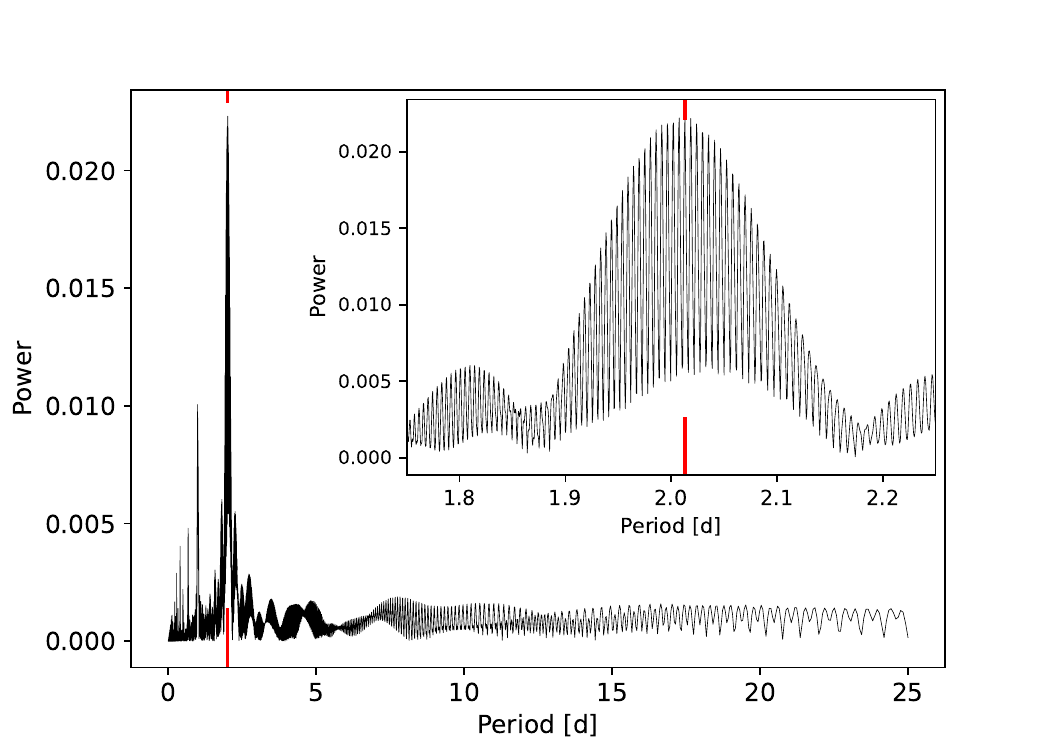}
\caption{\label{fig:ls} {  Lomb-Scargle periodogram of the out-of-eclipse TESS light curve. 
We measure the rotational modulation in the TESS light curve and obtain a periodicity of 2.013 $\pm$ 0.076~d, which is consistent with synchronous rotation. The most significant period is marked in both the plot and the inset with the vertical red lines. The inset shows the structure around the most significant peak, with its width determining the uncertainty in the rotation period. }}
\end{figure}

The EB MML~48 (TIC~129116176) was observed as part of a TESS Guest Investigator (GI) program (PI: Ricker; ID: G011280-G011154) in two-minute cadence in Sector~11 in 2019. Additionally, in Cycle 3 of the TESS GI program MML~48 was observed in Sector~38 (2021) in two-minute cadence (PI: Hebb; ID: G03143). In 2023, it was again observed in two-minute cadence in Sector 65. We downloaded the TESS light curves for all observed sectors from the{  Mikulski Archive for Space Telescopes} \footnote{https://mast.stsci.edu} using the SPOC data products. 

\begin{figure*}
\includegraphics[height=0.45\hsize,trim=0 6.5cm 0.5cm 6.7cm, clip]{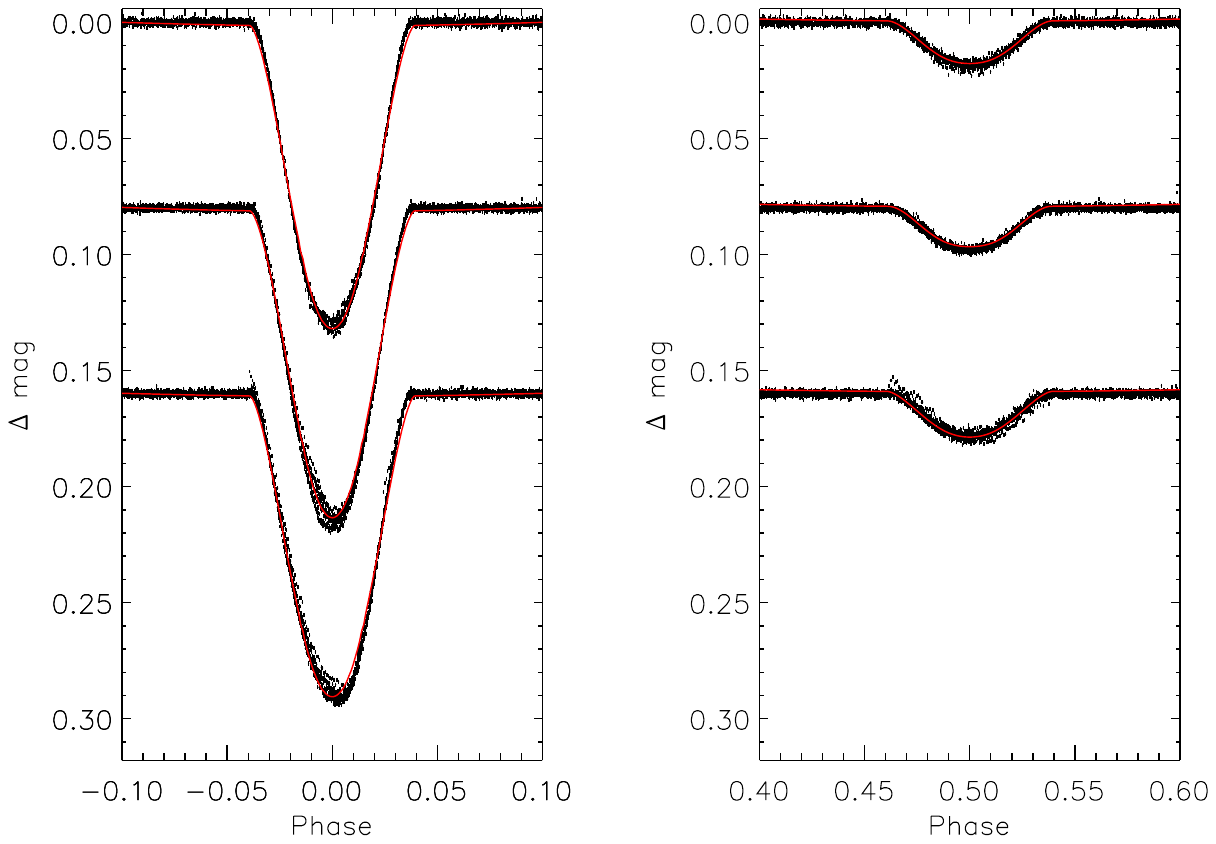}
\includegraphics[height=0.45\hsize,trim=0 6.5cm 8.7cm 6.7cm, clip]{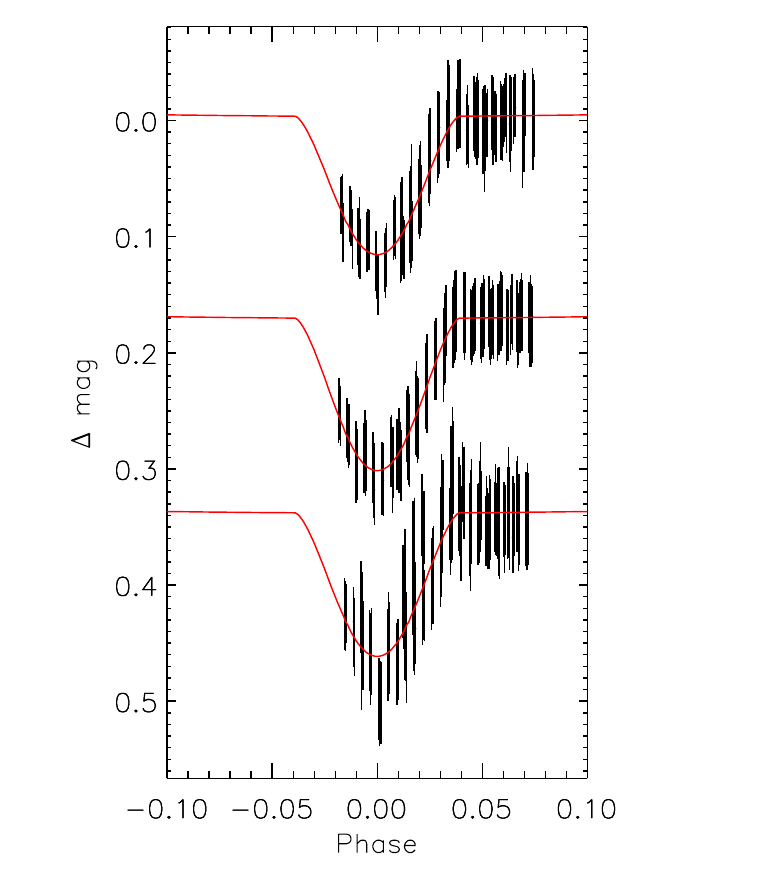}
\caption{TESS and CASLEO eclipse light curves of MML~48 compared to best fit model. Left: Primary-eclipse TESS light curve acquired during Sectors~11 (top), 38 (middle), and 65 (bottom). Overplotted is the {\sc Phoebe} best fit model in the I-band filter with a third light contamination (of 0.077 in units of total light for Sector~11,  0.064 for 38, and 0.091 for 65), and it is shown by the continuous red line. 
The individual data point photometric uncertainties are represented by the length of the vertical lines. 
 Center: Secondary-eclipse TESS light curve compared to the best fit model. The TESS light curves are plotted on the same scale for comparison. 
 Right: CASLEO light curves in VRI passbands (from top to bottom) of the MML~48 primary eclipse. 
}
\label{fig:lcstess}
\end{figure*}

The TESS light curves show several high-amplitude flares and out-of-eclipse variability, indicating clear spot evolution with a period consistent with synchronous rotation. 
Given that the peak-to-peak amplitude of the spot modulation varies between $\sim$3 and 10\% in relative flux, we attribute the measurable rotational modulation to the primary star. After masking the eclipses, we performed a Lomb-Scargle periodogram on the full TESS light curves deriving a rotation period of 2.013 $\pm$0.076~d (Fig.~\ref{fig:ls}). We used the python package {\sc Lightkurve} \citep{lightkurve}. The shape of the periodogram power distribution has a broad peak at 2.013 days (see the inset in Fig.~\ref{fig:ls}). It also shows high-frequency structure throughout, probably due to spot evolution on long timescales. The width of the highest peak, which indicates a range of rotation periods around the best period, is likely due to differential rotation. Thus, the uncertainty on the rotation period is derived from fitting the peak with a Gaussian and adopting the width of the Gaussian as the 1-$\sigma$ error bar. Finally, given that the orbit is circular and that the timescale for the tidal circularization is longer \citep[e.g.,][]{Zahn1989}  than the timescale for synchronization \citep[e.g.,][]{Hut1980, Zahn1977}, we expect both stars to be rotating synchronously with the orbital period. 
We removed the rotationally induced variability by fitting a polynomial to the out-of-eclipse data around each individual eclipse to rectify the light curve as shown in Figure~\ref{fig:lcstess}, but the shapes of the eclipses are still affected by the spots.  
A detailed spot model is beyond the scope of this paper, so we only used the TESS light curves for the ephemeris determination and not in the EB modeling of the components. 

\subsection{Spectroscopic data}

We acquired follow-up spectroscopic data that were used for stellar characterization and to measure {  RVs}. However, we note that we were not able to identify the secondary star in any of the spectroscopic datasets.  Below, we describe the spectra that we acquired, the corresponding reductions, and the measurements from each dataset. Table~\ref{tab:rvs} presents the primary sta RVs. 

\subsubsection{SOAR spectra} 
During the nights of 7 June, 2012 June, 2 July, 2012, 5 July, 2012, and 5 September, 2012,
MML~48 was observed with the Goodman High Throughput
Spectrograph \citep{Clemens2004} on the 4.1~m SOuthern
Astrophysical Research (SOAR) telescope at Cerro Pach\'on. The spectrograph was set up using the 1200~l/mm grating, a $0.46^{\prime\prime}$ wide slit, and a GG-495 blue-blocking filter to provide a wavelength range of $\sim$ 6275-7495\r{A}; it has a spectral dispersion of 0.304\AA\ per pixel on a 4096 x 4096 Fairchild CCD detector with 0.15-arcsec/pixel.  MML~48 was observed on five unique occasions over the course of these nights with 900s exposure times, resulting in a peak signal-to-noise ratio (S/N) of $\sim 300$ at 6500\AA. The nearby fainter star, 2MASS~14413595-4700280, was observed at the same epochs to determine its relationship to MML~48.

Standard \textsc{IRAF} procedures were used to process the raw images and then perform the trace, order extraction, and wavelength calibration of all spectra. \textsc{IRAF} was also used to measure the Heliocentric {  RVs}relative to the International Astronomical Union standard stars HR~6468, HR~6859, HD~120223, and HD~126053.  The measured RVs 
of MML~48 and their uncertainties derived from this analysis are given in Table~\ref{tab:rvs}.  

These measurements show MML 48 to have {  RV} variations in phase with the photometric period from the WASP light curves, confirming it as an eclipsing binary object. 2MASS~14413595-4700280 did not vary in velocity and is not gravitationally bound to the young system.

\subsubsection{FEROS  spectra} \label{sec:feros}

We obtained two 900~s exposures of MML~48 on the consecutive nights of 19-20 September, 2012 using the FEROS spectrograph ($R\sim 48000$) on the 2.2m MPG/ESO telescope at La Silla (PI: Faedi; ID: 089.C-0471).  These spectra were processed, extracted, wavelength calibrated, and continuum normalized using the \'echelle data-reduction package \textsc{REDUCE} \citep{piskunov2002} with calibration data obtained on the same night.  The final reduced spectra cover the wavelength range from $\lambda$ = 3527–9216$\AA$ and have a S/N $\sim 100$, as calculated in { 
\citet{Stoehr2008}} for a V$\sim 10^{\rm th}$ magnitude object such as MML~48.  

We applied our implementation of the least-squares deconvolution (LSD)
technique \citep{Donati1997} to the final FEROS spectra to obtain very high S/N ($\sim 230$) average absorption-line profiles of the primary component of MML~48 on each night.  We fit a G5-type broadened synthetic spectrum using the VALD line list as incorporated in {  Spectroscopy Made Easy \citep[SME;][]{Valenti1996}} 
The synthetic spectrum was convolved with a Gaussian instrumental broadening of 2 \kms\ and an additional rotational broadening profile.  Despite the extremely high S/N, the absorption-line profile of the secondary star was not visible in these data, suggesting a very-low-luminosity secondary component.   However, this analysis did allow us to measure, for the primary component, both its RV (reported in Table~\ref{tab:rvs}) and its \vsini\ of 42~\kms.  
We adopted an uncertainty of 1.0~\kms\ in the RV measurements to account for visible asymmetries in the LSD profile, presumably due to 
{  activity and/or starspots}.

\begin{table}
\caption{Radial-velocity measurements for {  p}rimary EB component.}\label{tab:rvs}
\begin{tabular}{c c c c c c}
\hline\hline\\
 BJD$_{\rm TDB}$      & RV$_{\rm prim}$      & $\sigma_{\rm RV_{\rm prim}}$      & Instrument   \\
 $- 2\,450\,000$ & (kms$^{-1}$) & (kms$^{-1}$) & \phantom{Instrument} \\
\hline\\
           6086.5359 &       21.20 &        2.00 &  SOAR \\ 
           6111.5201 &      -26.30 &        2.00 &  SOAR \\ 
... & ... & ... & ... \\
\hline\\
\end{tabular} \\
{\footnotesize The full table is available at the CDS.}
\end{table}

\subsubsection{CTIO/SMARTS CHIRON spectra} 
Eleven spectra of MML~48 were taken in fiber mode {  (R$\sim24\,700$)} with the CHIRON \'echelle spectrograph \citep{Schwab2012,Tokovinin2013} on the CTIO/SMARTS 1.5 m telescope.   
Ten spectra were taken between 16 and 21 March, 2019 to be coincident with the follow-up time-series photometry taken on the CTIO/SMARTS-0.9 m (Sect.~\ref{sec:ctiophot}), and an additional single spectrum was obtained on 28~May,~2019.
Standard data processing of the raw science and calibration images along with the trace, order extraction, and wavelength calibration of the spectra were completed with custom \textsc{IDL} programs.  This process is described in detail in \citet{Walter2018}.
{  The RV} for the primary star was derived from the final CHIRON spectrum at each epoch by applying the same LSD deconvolution analysis described in Sect.~\ref{sec:feros} that we applied to the FEROS spectra.    We adopted a typical uncertainty of 0.5~\kms\ following previous experience with the high-stability CHIRON spectrograph.   Again, we failed to identify the absorption-line profile of the secondary star in these optical wavelength spectra, further confirming its very low luminosity.  Furthermore, given the lower resolution of our CTIO data as compared to the FEROS data, we did not measure the \vsini.

\section{Analysis and results} \label{sec:analysis}

We present the analysis used to derive fundamental properties of MML~48 given all the available data. The process{  toward} our preferred results is iterative, ensuring that the final fit was consistent throughout our datasets and analyses. The adopted physical properties of the MML~48 system and its stellar components are presented in Table~\ref{table:eb}. 

\subsection{Orbital period and primary-eclipse epochs from light curves}\label{sec:timings} 

In this section, we describe the analysis required to derive the eclipsing binary period (P$_{\rm orb}$) and time of primary eclipse ($T_0$). The final values resulting from the simultaneous fit to the highest precision photometry available are reported in the first section of Table~\ref{table:eb}. 

We restricted our period analysis to the highest precision light curves, namely the FTS secondary eclipse from 2013, the CTIO March 2019 primary eclipses, and both primary and secondary eclipses in the three TESS sectors (11, 38, and 65).  We fit them using the dedicated EB code {\sc Phoebe} \citep{Prsa2005}, which allows for a simultaneous fit to light curves in different pass bands. 
We chose the epoch of the first primary eclipse of the March CTIO data as it was relatively close in time to most of the data used in the fit.  
Although we did not use the TESS light curves to derive radii of the eclipsing stars, we were able to use these data for the ephemeris derivation. We used the solution from Sect.~\ref{sec:phoebe} that was already a good fit to the light curves and incorporated the TESS light curves using the Johnson I pass band and the third light to account for the close visual companion $\sim$15\arcsec\ away, 2MASS~14413595-4700280, and we fit the limb-darkening coefficients to best fit the eclipses. We thus obtained a good fit, as shown in Figs.~\ref{fig:lcsfit} and \ref{fig:lcstess}. In letting the period and epoch be free parameters, we were then able to derive the ephemeris reported in Table~\ref{table:eb} from the simultaneous best fit to the most precise light curves from 2013 to 2019. The errors reported on both the period and epoch are the formal uncertainties from this best fit.

\begin{figure}
   \centering
    \includegraphics[width=\columnwidth,trim={0.7cm 6.7cm 0 6.7cm},clip]{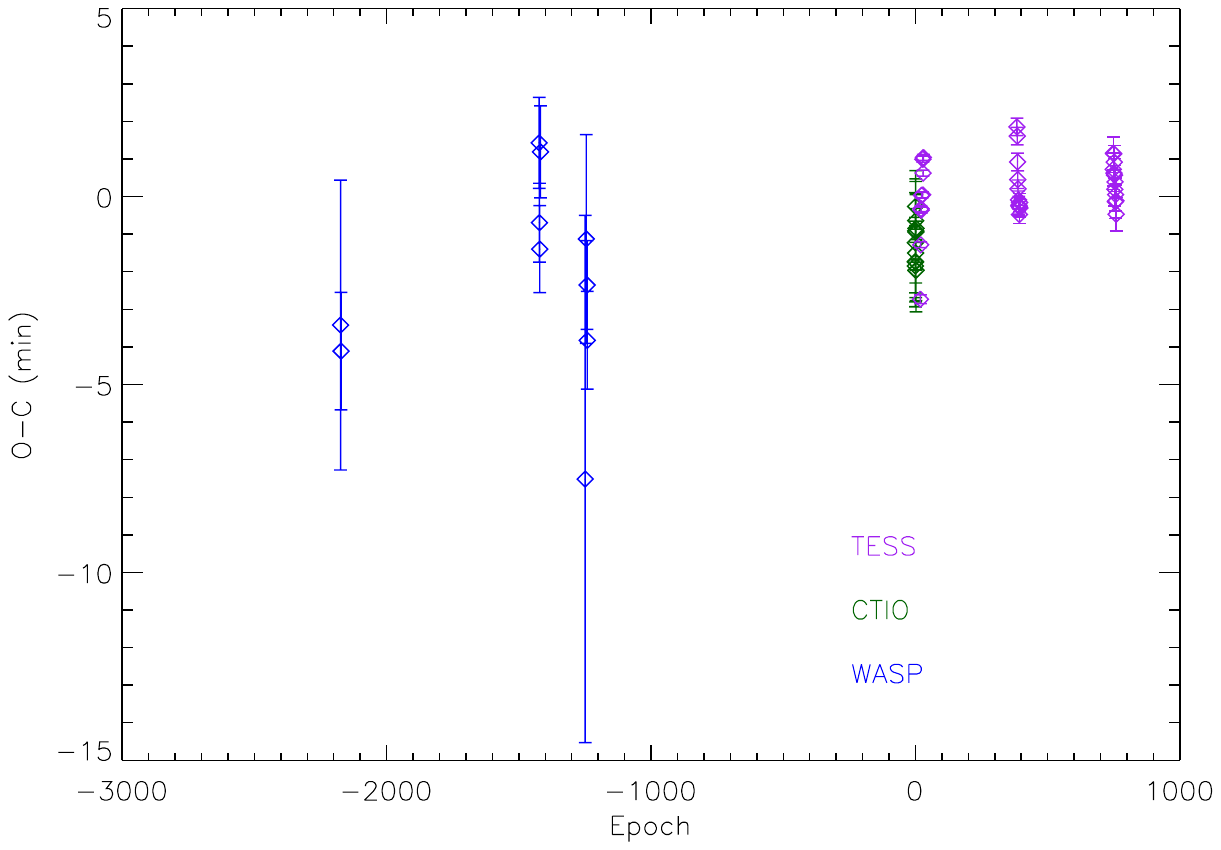}
   \caption{O-C diagram for all full primary eclipses spanning from 2007 to 2023.  The O-Cs were calculated on the measured time of minimum for each primary eclipse where both sides of the out-of-eclipse data are available (Table~\ref{tab:epochs}) and the ephemeris derived from the highest precision data as described in Sect.~\ref{sec:timings}. The WASP primary eclipses are shown in blue, the CTIO primary eclipses are shown in green, and the TESS primary eclipses are shown in purple. There is no observed trend in the O-C diagram, and all the data are consistent within their uncertainties with the derived linear ephemeris. It must be noted that there is a spread of a few minutes in the individual epoch measurements from a single season that we attribute to spots and that caused the initial uncertainty in the period determination.} 
    \label{fig:oc}
    \end{figure}

Initially, we used the discovery WASP data to derive the EB ephemeris using the same procedure as described in \citet{GomezMaqueoChew2019}, both by measuring an average T$_0$ for each WASP season and by measuring individual T$_0$ for every full primary eclipse in the data by fitting an EBOP model \citep{southworth2007}. 
We utilized the WASP data encompassing from 2006 to 2014. Unlike with MML~53, we did not identify light-time variations in the individual T$_0$ measurements.  As the CTIO and TESS data were obtained, we derived ephemerides from each full sector and T$_0$ values for every individual full primary eclipse using rectified light curves. In the case of the precise TESS light curves, we simultaneously fit a sine curve and the EB model to account for short period variations during the eclipses. 
The individual epochs for the full primary eclipses derived from this analysis are presented in Table~\ref{tab:epochs}, and the O--C plot with the best ephemeris is shown in Fig.~\ref{fig:oc}. 

The long baseline of the data should provide an extremely precise constraint on the period of the EB, but the copious spot modulation in and out of the eclipses causes significant variations in the eclipse timings.  
It is possible that the epoch of an individual eclipse is skewed by a star spot, and the epoch of an entire dataset could be modified by a consistent spot \citep[e.g.,][]{Watson2004}. It is clear from Fig.~\ref{fig:oc} that the scatter in eclipse timings is much greater than the error on the individual measurements, particularly for the extremely precise photometry obtained with TESS.
Consequently, the period derived from eight years of WASP data was inconsistent with the newer TESS data.
The epoch measured from 2019 data is earlier
by eight minutes than what would be predicted by propagating the ephemeris from the earlier WASP data.  Believing the uncertainty on the WASP data, the difference in epoch ten years later should be at most 78s. Similar discrepancies were evident when comparing the timing of the WASP data with the ephemeris derived only from the TESS Sector~11 light curve.

\begin{table}
\caption{Time of minimum for full primary eclipses of MML~48.}\label{tab:epochs}
\begin{tabular}{l l c }
\hline\hline\\
Dataset & Filter & Epoch (BJD$_{\rm TDB}$- 2\,450\,000) \\
\hline\\ 
WASP 2007   &        WASP   & 4176.52237 $\pm$ 0.00253 \\
WASP 2007   &        WASP   & 4180.55610 $\pm$ 0.00065 \\
... & ... & ...  \\
\hline\\
\end{tabular} \\
\footnotesize{The full table is available at the CDS.}
\end{table}

We explored whether an eccentric orbit would have an effect on the derived ephemeris, by letting the period, epoch, eccentricity $e$, angle of periastron $\omega,$ and phase shift be free parameters. Within the uncertainties, we obtain consistent period and epoch values. However, the fit with a very small eccentricity  0.0006 has a slightly larger $\chi^2$ than the circular solution. With this, we conclude that the orbit is likely not eccentric and that the ephemeris using a circular orbit is robust.

\begin{table*}
\caption{Physical properties of MML~48 eclipsing stars and their orbit.}             
\label{table:eb}      
\centering          
\begin{tabular}{l c c c }    
\hline\hline       
\multicolumn{2}{c}{Parameter} & Value & Units\\ 
\hline               
Orbital period & \porb  &  2.0171068  $\pm$  0.0000004 & days\\ 
Time of primary minimum & T$_0$ & $2458561.71492 \pm 0.00003$ &   BJD$_{\rm TDB}$ \\  
\hline
Primary temperature & \teffp & 5386 $\pm$ 100 & K\\ 
Primary rotational velocity & \vsini & 42.0 $\pm$ 2.0 & \kms \\ 
Primary mass & \mprim & 1.2  $\pm$  0.07 & \msun \\
\hline
Eccentricity & $e$ & 0. & (fixed)  \\ 
RV semi-amplitude & $K_1$ & 32.86 $\pm$ 0.33 & \kms \\
CHIRON Systemic velocity  & $\gamma_{\rm CHIRON}$  & 5.6 $\pm$ 0.15 & \kms \\
FEROS Systemic velocity  & $\gamma_{\rm FEROS}$  & 6.2 $\pm$ 0.7 & \kms \\
SOAR Systemic velocity  & $\gamma_{\rm SOAR}$  &  4.8 $\pm$ 0.9 & \kms \\
\hline
Mass ratio & $q_{\rm EB}$ = \msec/\mprim &  0.209  $\pm$  0.014 & \\
{Semi-major} axis & $a$ & 7.61  $\pm$  0.12 & \rsun\\ 
 & $a$ & 0.0354  $\pm$  0.0006 & au \\  
 Secondary mass & \msec &  0.2509 $\pm$  0.0078  & \msun \\
 \hline
Inclination & $i$ &     81.372  $\pm$  0.045  & \degree \\
Primary potential & $\Omega_1$ &   5.057  $\pm$  0.013 & \\
Secondary potential & $\Omega_2$ &    4.100  $\pm$  0.013 & \\
Temperature ratio & \teffs/\teffp &  0.642  $\pm$  0.001 & \\
Primary radius & \rprim &  1.574  $\pm$  0.026  $\pm$ 0.050 & \rsun \\ 
Primary surface gravity & \loggp &    4.12  $\pm$  0.03 & dex (cgs) \\ %
Secondary radius & \rsec & 0.587  $\pm$  0.0095 $\pm$ 0.050 & \rsun \\ 
Secondary surface gravity & \loggs &  4.28  $\pm$  0.02 & dex (cgs) \\
Secondary temperature & \teffs & 3456 $\pm$  100 & K \\
\hline       
\end{tabular}

\end{table*} 

\subsection{Primary effective temperature and EB metallicity from FEROS spectra}\label{sec:starchar} 

We modeled the highest S/N FEROS spectrum of MML~48 with SME   \citep{Valenti1996}  using the VALD3 line list to derive the stellar parameters as described in \citet{GomezMaqueoChew2019}.  
The FEROS spectra have the highest resolution of all of our spectroscopic data, so they provide the best constraint on the stellar parameters.  In addition, this spectrum is single lined, showing only flux from the primary star given the very   large luminosity ratio in the spectrum.   The SME spectral synthesis gives a best fitting effective temperature for the primary star of \teffp=$5386 \pm 100$~K assuming solar metallicity for this young system.   We adopted a fixed log~g based on the mass and radius measurements derived from the EB analysis. At \teff=5300~K, the Mg~I triplet starts to become too deep, while for \teff$=5500$~K the Na D lines start to be too shallow, indicating a temperature between these two values. In addition, both of these lines are sensitive to log~g, and the width indicates that the log~g derived from the mass and radius is very reasonable for this star.   
The SME synthesis also gives \vsini$= 42.0$\kms.   Finally, there appears to be some filling in of the H-$\alpha$ line, which is not uncommon for a young, active star.

\subsection{RV semi-amplitude  and systemic velocity from SB1 model}\label{sec:sb1} 
\begin{figure}
\centering
\includegraphics[width=1.0\hsize]{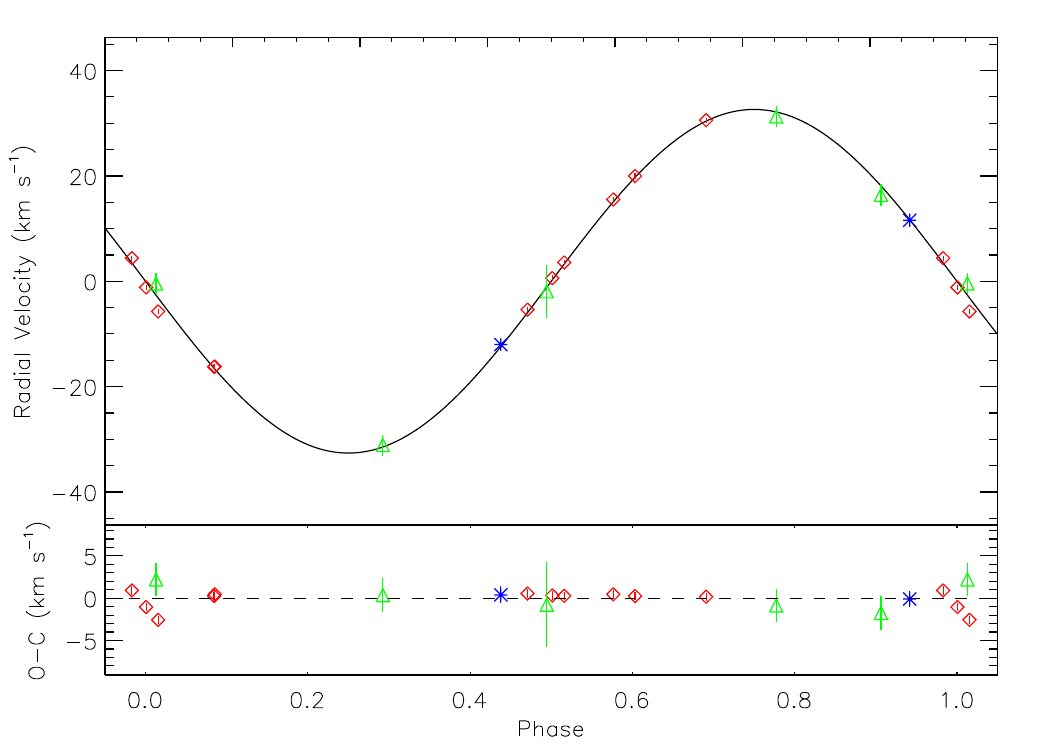}
\caption{Relative {  RV} curve for MML~48. Top panel: We present the three RV datasets CHIRON (red diamonds), FEROS (blue asterisks), and SOAR (green triangles). Each dataset has been shifted by the corresponding systemic velocity (reported in Table~\ref{table:eb}).  Overplotted with a solid black line is the model RV curve fit to all three datasets, as described in Sect.~\ref{sec:sb1}. Bottom panel: Residuals to RV model shown above for all RV measurements. The errors of the individual RV measurements are shown and may be smaller than the{  data point}.}
\label{fig:rvs}
\end{figure}

Once the binary ephemeris was determined (Sect.~\ref{sec:timings}), we combined all three RV datasets to derive the RV-dependent properties. In particular, we derived the semi-amplitude of the RV curve ($K_1$) and the systemic velocity of each dataset ($\gamma_{\rm CHIRON}$, $\gamma_{\rm SOAR}$, and $\gamma_{\rm FEROS}$) to account for zero-point differences between the instruments. For this analysis, we utilized \texttt{RadVel}, a python package that models the Keplerian RV motion and that was developed for transiting planets \citep{Fulton2018}, and it is appropriate for this system given the single-lined nature of MML~48.  We assumed a circular orbit and a constant systemic velocity (i.e., $\gamma$-acceleration terms being zero). We set the jitter term to zero, as the uncertainty in the RV measurements is on the order of 1~\kms\ and already encompasses this source of RV noise. We fit for $\log K_1$ and the three $\gamma$s using the \texttt{RadVel} Markov chain Monte Carlo capability with 50 walkers, 
discarding the initial 420\,000 steps in the burn-in phase and until both converging conditions were attained (after 880\,000 steps in this case); this was done as described in \citet{Fulton2018}. The measured parameters are reported in the third section of  Table~\ref{table:eb}, with their corresponding 1-$\sigma$ uncertainties and the best fit model is shown with the RVs in Fig.~\ref{fig:rvs}. 

Additionally, to assess whether our choice of a circular orbit is appropriate, we fit the RV data allowing for the semi-amplitude ($K_1$), the three systemic velocities ($\gamma_{\rm CHIRON}$, $\gamma_{\rm SOAR,}$ and $\gamma_{\rm FEROS}$), eccentricity $e,$ and angle of periastron $\omega$ to be free parameters. All derived parameters were consistent with those of the circular orbit solution within 1$-\sigma$, and we obtain a small eccentricity (e = 0.010 $\pm$ 0.008), which has an $\sim$40\% probability of being spurious \citep{Lucy1971}. It must be noted that this Lucy-Sweeney probability is a lower limit, as the light curves are not included in this assessment, but they also contribute to constraining $e$ and $\omega$ via the time and duration of the eclipses \citep{Kallrath2009}. Moreover, in Sect.~\ref{sec:phoebe}, we find no evidence of eccentricity in the light curves. Thus, we assume a circular orbit.

\subsection{Semi-major axis, mass ratio, and secondary mass from Keplerian motion} \label{sec:kepler} 
Based on our measurements of the semi-amplitude of the RV curve (K$_1$), the orbital period (P$_{\rm orb}$), and assuming a circular orbit, we derived the semi-major  axis (a) and M$_2$ as a function of M$_1$ by inverting the equation for the mass function \citep[Eq.~4.4.34 in][]{Kallrath2009}. We utilized the nominal values for the solar constants following the IAU 2015 Resolution B3 \citep{Prsa2016}.  
Using the derived primary mass (Sect.~\ref{sec:m1}), we calculated a mass ratio (q = M$_2$/M$_1$). The uncertainties in the secondary mass are derived from propagation of error analysis, including the uncertainty in K$_1$, P$_{\rm orb,}$ and M$_1$. These values are reported in the fourth section of Table~\ref{table:eb}.

\subsection{Primary mass from PMS evolutionary models}\label{sec:m1}  

\begin{figure}
\centering
\includegraphics[width=1.0\hsize]{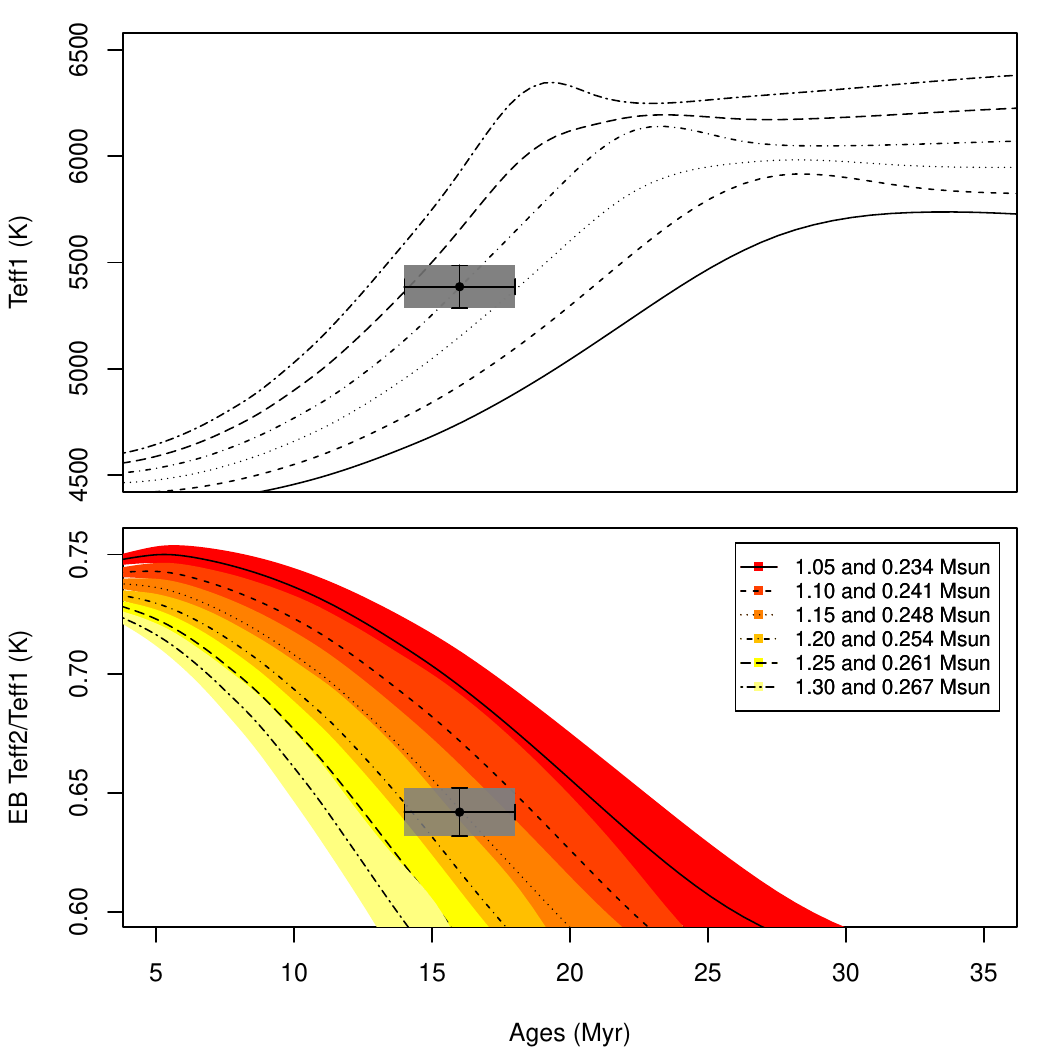}
\caption{Estimation of mass of primary  star done by comparing the \citet{Baraffe2015} stellar-evolution models to measured effective temperature of primary star (Sect.~\ref{sec:feros}) in the top panel and to temperature ratio measured from the relative eclipse depths (Sect.~\ref{sec:phoebe}) in the bottom panel at the measured age of 16 $\pm$ 2 Myr for UCL.  Our measurements are marked by the black point with a 1-$\sigma$ uncertainty given in the gray-shaded region. The different lines show the evolution of stars with masses between 1.05 and 1.30 \msun; the color-shaded areas show the range of temperature ratios associated with each primary star mass.}\label{fig:m1}
\end{figure}

All the optical spectra presented in this paper show only the spectrum of the primary star due to the very{  large} luminosity ratio between the two EB components.  Thus, the masses of both components cannot be directly measured in this single-lined system, and we must derive the primary star's mass independently from stellar evolution models.  This is a common technique in the analysis of transiting planets, which are effectively single-lined eclipsing systems with very{  large} luminosity ratios.  In the case of a planet transiting a quiescent star, the effective temperature (and metallicity) of the primary combined with the mean stellar density derived from the shape of the light curve constrain the mass of the primary star \citep[e.g.,][]{Enoch2010,Hebb2009,Hebb2010b}.  

In the case of young stars that are known to be larger than those predicted by standard evolutionary models, this technique provides an inaccurate estimate of the stellar mass. 
Indeed, for the young, double-lined EB MML~53, where the masses were directly measured \citep{GomezMaqueoChew2019}, we found that comparing the radii of the component stars to stellar evolution models led to a significant underestimation of the age of the system and an overestimation of the stellar masses that would have been derived from these same models. Instead, we found that the masses derived from standard stellar models based on the temperature ratio of the binary components of MML~53 agreed with the measured masses.  Thus, we compared the independent measurements of the effective temperature of the primary star and the EB temperature ratio to the standard models to estimate the mass of the primary star in MML~48.   

In the top panel of Fig.~\ref{fig:m1}, we show the age and primary star temperature of MML~48 compared to stellar-evolution models from {  \citet{Baraffe2015}} with a range of masses from $1.05-1.30$~\msun.  We adopt the age of $16\pm 2$~Myr as derived for Upper Centaurus Lupus \citep{Mamajek2002, Pecaut2012, Pecaut2016,Wright2018} and the temperature of the primary star derived from the spectra (Sect.~\ref{sec:feros}).  

In the bottom panel, we compare the temperature ratio derived from the multiband light curves (Sect.~\ref{sec:phoebe}) to these same models because it is a completely independent constraint on the temperatures of the stars.  As described in Sect.~\ref{sec:kepler},  each primary star mass determines a unique \msec\ and mass ratio ($q$) from the measured K$_1$ and \porb.  The evolution of the model temperature ratio for each black line corresponds to the temperature derived from the mass tracks interpolated at the \msec\ value divided by the temperature of the primary star from the \mprim\ track.  The color band around the line reflects the range of temperature ratio values from the tracks that are possible within the uncertainties of the measured quantities.
 The two independently measured properties overlap with the mass tracks from $1.15-1.25$\msun, so we adopted a mass for the primary star of $1.2\pm 0.07$\msun, taking a conservative estimate for the uncertainty.

\subsection{Radii, temperature ratio, and inclination from EB model}\label{sec:phoebe} 

Using the information derived in the previous sections, we modeled the FTS (Sect.~\ref{sec:fts}) and CTIO/SMARTS 0.9m March 2019 (Sect.~\ref{sec:ctiophot}) light curves using the specialized EB code {\sc Phoebe} \citep{Prsa2005}. 
The fit light curves, best fit EB model described in this section, and residuals to the fit are shown in Fig.~\ref{fig:lcsfit}. The resulting derived parameters are in the last section of Table~\ref{table:eb}. 

For the best fit model, we varied the potentials of the eclipsing stars ($\Omega_1, \Omega_2$), the inclination angle ($i$), and the effective temperature of the secondary (\teffs). Following \citet{Torres2021}, we adopted the bolometric gravity-darkening exponents from Fig.~1 of \citet{Claret2000}. For the primary component with a $\log{T_{\rm eff,1}} = 3.7, $ we used a $\beta_1$=0.40, and for a the secondary one with a $\log{T_{\rm eff,2}} = 3.5, $ we used a $\beta_2$=0.20.
A reflection albedo of 0.5 for each component was adopted, as is appropriate for stars with convective envelopes \citep{Rucinski1969}. To assess the effect of the choice of albedo on the EB model, we compared a model with an albedo of 0.0 for both stars, with 1.0 and with our nominal value of 0.5.  We find that the effect of the albedo between our model and these two models is smaller than the standard deviation of the residuals in every filter. 
We assume both components are rotating synchronously to the orbital period. Given the results above, we adopted a circular orbit.  
The fit was done with an Levenberg–Marquardt algorithm, internal to {\sc Phoebe}, which minimizes the cost function to all the light curves simultaneously.  At each step, the luminosities for each pass band were calculated, and the limb-darkening coefficients were interpolated. We started the fit multiple times from different initial parameters that were found by hand, ensuring that the converged solution was not a local minimum. 
The uncertainties of the fit parameters were obtained from the covariance matrix and are the formal errors. Other derived properties were calculated using standard constants \citep{Prsa2016}, and their uncertainties were derived from error propagation analysis. All fit and derived parameters and the corresponding uncertainties are reported in the bottom section of Table~\ref{table:eb}.  We estimated the relative flux between the components in the different filters from the light-curve model to be of 0.0032 in B, 0.0060 in V, 0.0092 in R, 
0.0154 in I, and 0.0243 in z$^\prime$. 

Given the equal duration of the primary and secondary eclipses and their separation in phase \citep{Kallrath2009}, we confirmed, using the light-curve data, that the orbit is most likely circular. We additionally tested whether a nonzero eccentricity could be fit to the light curves (also fitting for the angle of periastron and the phase shift, as is appropriate in {\sc Phoebe}), and, unsurprisingly, we were unable to achieve a solution that converged. We thus consider that our initial assumption that the orbit is circular is correct, which is to be expected given the short period of the EB and its age.  

In Fig.~\ref{fig:lcstess}, we show the comparison between the best fit EB model and TESS (Sect.~\ref{sec:tess}) and the CASLEO light curves (Sect.~\ref{sec:casleo}). 
The TESS pass band is not included in the version of {\sc Phoebe} that we used; instead we used an I-band filter for the TESS {\sc Phoebe} model, which is similar to the TESS response function. 
We fit for the third light for the TESS light curves, obtaining a value of 7.7\% of the total light of the system for Sector~11, 6.4\% for 38, and 9.1\% for 65, which are roughly consistent with the contamination factor from the TIC v8.2 \citep[0.1577;][]{Paegert2021}.
We can see in the left and center panels of Fig.~\ref{fig:lcstess} that this best fit model  matches relatively well to the rectified light curves, but there are still residual spot modulations. The CASLEO light curves were not included in the fit because the significant spot modulation present could not be removed robustly; this is due to the fact that the eclipses are partial and have relatively lower photometric precision than the CTIO light curves.
Including the TESS and CASLEO light curves would introduce systematic uncertainties in the derived physical properties of the MML~48 eclipsing stars, and thus we decided not to use these two datasets in the EB fit. They are only shown for completeness. A future analysis of the TESS light curve is planned and these limitations will be accounted for.

To check the radius of the primary star derived from the EB model, we calculated the radius assuming that the stellar rotation is synchronized with the orbital motion and using the measured \vsini\ of the primary star (42 $\pm$ 2 \kms; Sect.~\ref{sec:feros}).  We derive a radius for the primary star of 1.694 $\pm$ 0.081 \rsun. We also carried out a single-star SED fit to the broadband photometry of MML~48 and derive a primary star radius of 1.458 $\pm$ 0.054 \rsun, with an A$_V$ = 0.33 $\pm$ 0.04 and 
F$_{bol}$ = 3.965 $\pm$ 0.046 $\times 10^{-9}$  erg s$^{-1}$ cm$^{-2}$.  Therefore, we added a $\pm 0.05$ systematic uncertainty to the primary and secondary star radii to the formal errors from the EB analysis so that all estimates of the primary star radius are consistent within their 1-$\sigma$ uncertainties.

\begin{figure}
\centering
\includegraphics[width=1.0\columnwidth,trim={0.7cm 6.6cm 0 7.1cm},clip]{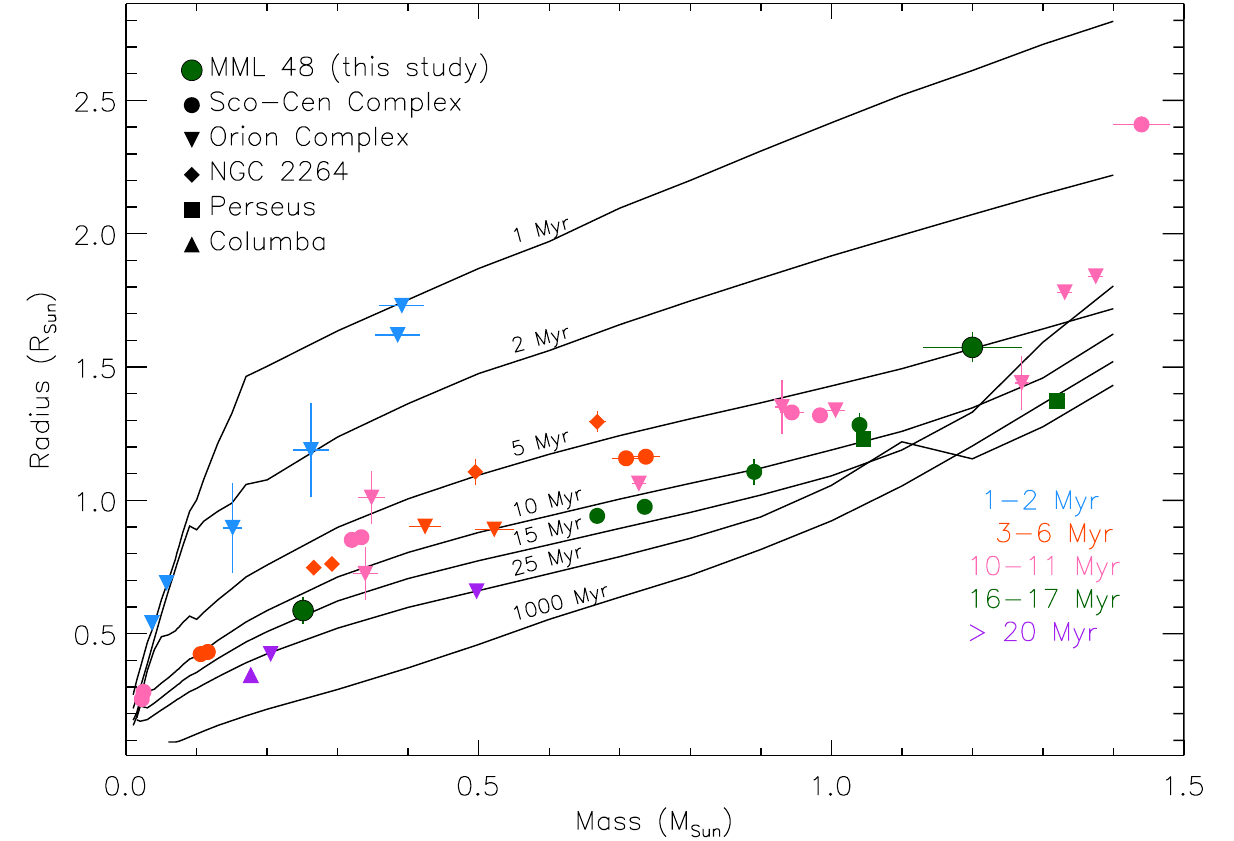}
\caption{Mass-radius diagram of  
known sample of pre-main-sequence EBs compared to low-mass stellar models from \citet{Baraffe2015}. MML~48 is shown in the large, green dots with uncertainties from the best fit solution. Also in green are the other eclipsing stars with age estimations between 16 and 17 Myr; pre-main-sequence EBs with ages from 1-2 Myr are in blue, 3-6 Myr in orange, 10-11 Myr in pink, and more than 20 Myr in purple. The different points show EBs in the five star-forming regions; EBs in Sco-Cen are shown by dots, in Orion by downward-pointing triangles, in NGC~2264 by diamonds, in Perseus by squares, and in Columba by  upward-pointing triangles. In general, the EB stars follow the behavior of the stellar tracks at larger radii, as expected for young stars and standard models. The radius of MML~48~A is larger than the radius of NP~Per~A by $\sim$15\%, as MML-48~A builds up $^3$He in its core (see also Figs.~\ref{fig:cno_analysis}--\ref{fig:cnobump}). The errors in all the measured masses and radii are included and in some cases are smaller than the points. 
}
\label{fig:mr}
\end{figure}

\section{Discussion}
\label{sec:discussion}

The EB MML~48 joins the short list of intermediate-age EBs composed of stars that are still in the pre-main sequence and can be used to constrain stellar evolution during a time when they are quickly evolving and arriving onto the main sequence.
We note that although many young EBs are in {  higher order} 
multiple systems \citep{Stassun2014},  we find no definitive evidence of a third star in the spectra, the {  RVs}, or the timing of the eclipses. 

It is typical to assume that that the stellar components of a close binary system are formed at the same time and are coeval; thus, the components can be described by a single isochrone.  In the case of the MML~48 stars, the very{  high} mass ratio between the eclipsing stars is able to constrain a single isochrone more tightly than two stars with more similar masses. 
All other known young EB systems have mass ratios that are higher than 0.5, whereas MML~48 has one of $q = 0.209$.  However, because of this difference in mass and despite using a least-squares deconvolution analysis (Sect.~\ref{sec:feros}) to create an extremely high S/N line profile from the highest resolution spectra, we were unable to extract the {  RV} of the secondary star due to its low relative luminosity compared to the primary. The secondary star is still below the detection threshold in our spectra; thus, we were unable to unambiguously measure the model-independent masses of both stars in this system. As such, the measurements presented in this discovery paper are based on a single-lined eclipsing binary system. 

For MML~48 to truly constrain stellar evolution, direct measurements of the fundamental properties of the MML~48 stars based on a double-lined eclipsing system are needed. We consider this to be beyond the scope of this discovery paper. Below, we compare our measurements for MML~48 to other known, young EB stars and to both standard and magnetic evolution models to put the system into the context of what is currently known of young, low-mass stars. 

As shown in Fig.~\ref{fig:mr},  the radius of MML~48~A is larger than the radius predicted by the 15 Myr isochrone, and it is approximately 15\% larger than the radius of NP Per A. It is surprising that even though they are the same age, the more massive NP Per A has a smaller radius than MML~48~A. 
Figure~\ref{fig:mr} does provide a hint to a possible mechanism that could lead to MML~48~A having a larger radius than other stars of similar masses. The 15 Myr and 25 Myr isochrones show an increase in radius for stars around $1.0~M_\odot$. This ``fusion bump" in the mass-radius diagram is the result of nuclear fusion processes in the stellar core as $^3$He is evolving toward its equilibrium abundance set by the {  proton-proton (p-p)} I chain \citep{Chabrier1997,Stassun2014}.

\begin{figure}
\centering
\includegraphics[width=1.0\hsize]{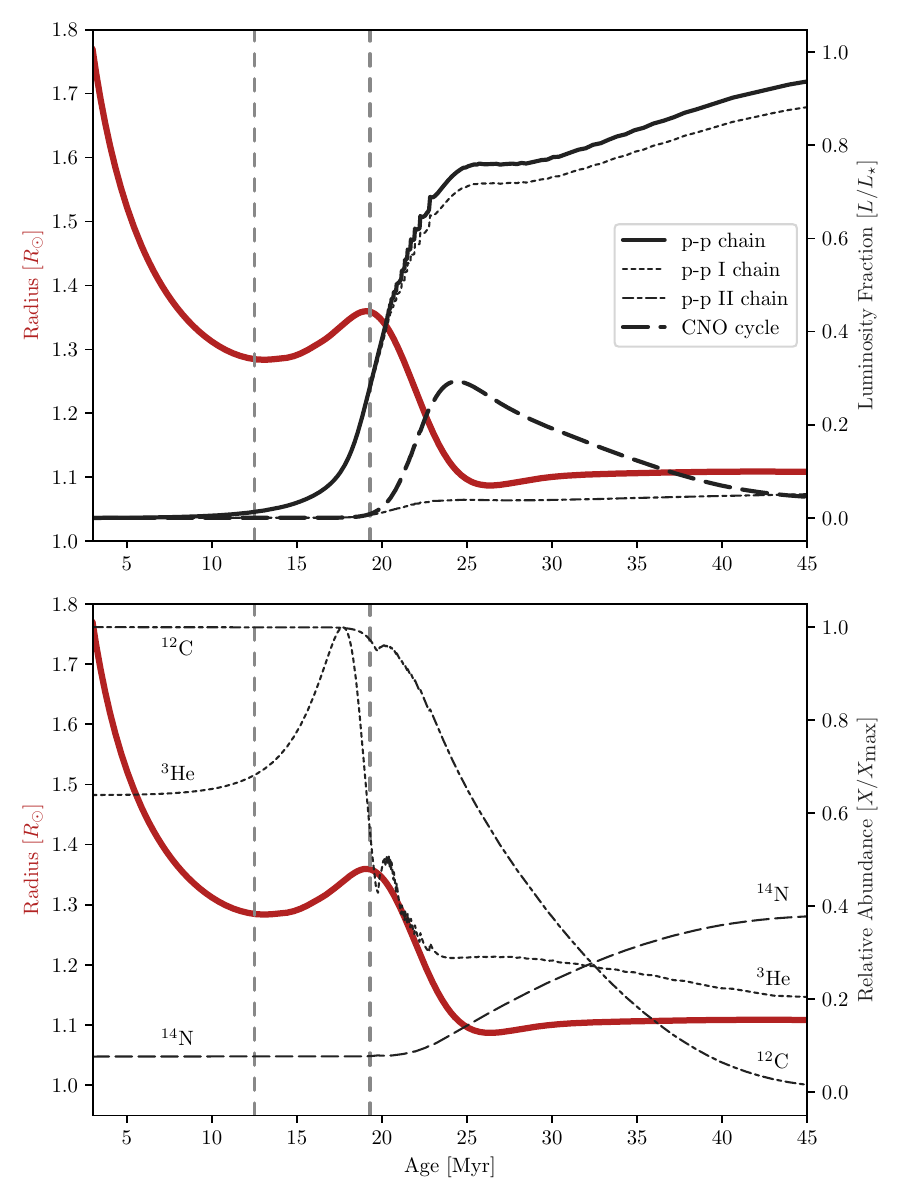}
\caption{Radius evolution of a $1.2~M_\odot$ nonmagnetic Dartmouth stellar-evolution model mass track (solid red line). This is shown with (top) the luminosity evolution of the p-p chain (solid black line) and CNO cycle (dashed black line) as a fraction of the total stellar luminosity and (bottom) the evolution of the central abundance of $^3$He, $^{12}$C, and $^{14}$N. Vertical lines denote the start and end of the first fusion bump.}
\label{fig:cno_analysis}
\end{figure}

\begin{figure}
\centering
\includegraphics[width=0.95\hsize,trim={0.7cm 0.2cm 0.2cm 1cm},clip]{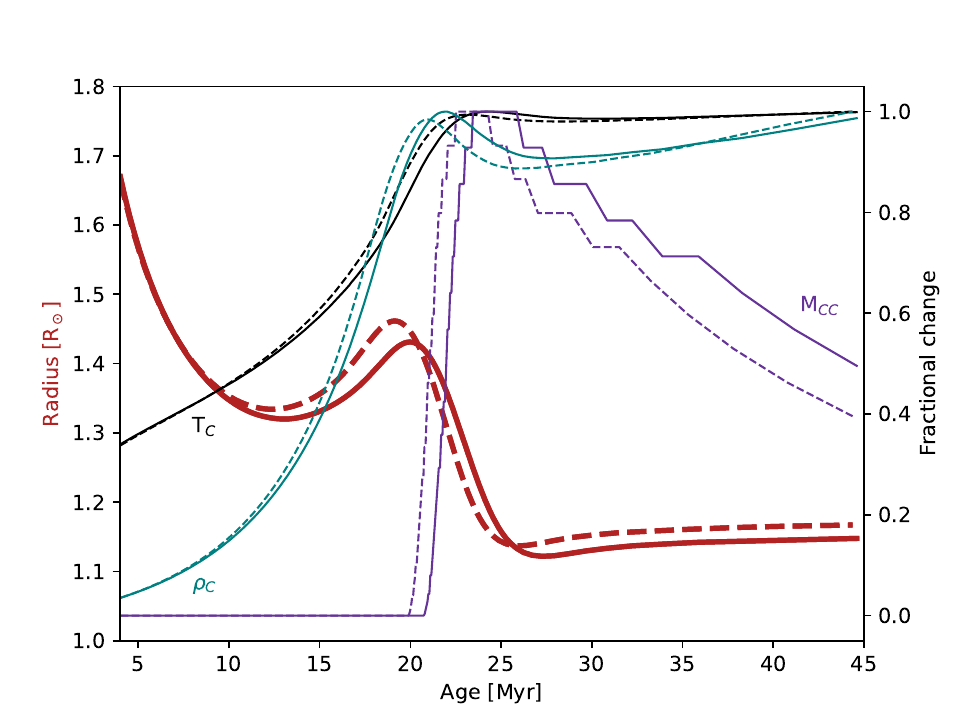}
\caption{Radius evolution of a $1.2~M_\odot$ stellar evolution model mass track (in red) from
\citet{Baraffe2015} for 
Y = 0.28 (solid lines) and Y = 0.29 (dashed lines). Y is the initial {  helium mass} fraction. 
This is shown with the evolution of the stellar central temperature ($T_C$; black),   the stellar central density ($\rho_C$; cyan), and the mass of the convective core ($M_{\rm CC}$; purple) as a fraction of the maximum value attained between 4 and 45 Myr  (see Sect.~\ref{sec:discussion}).  In the case of Y = 0.28 (solid lines), these maximum values are 
$T_C$ = 1.530$\times 10^7$ K, 
$\rho_C$ = 78.596 g cm$^{-3}$, and
 $M_{\rm CC}$ = 0.1023 \msun; in the case of Y = 0.29 (dashed lines), these are
 $T_C = 1.555\times 10^7$ K, 
$\rho_C$ = 80.575 g cm$^{-3}$, and
 $M_{\rm CC}$ = 0.1097 \msun. }
\label{fig:baraffemasstrack}
\end{figure}

Figure~\ref{fig:cno_analysis} shows the predicted radius evolution for a $1.2~M_\odot$ star from{  nonmagnetic} Dartmouth stellar-evolution models \citep{Feiden2016}. Similarly, Fig.~\ref{fig:baraffemasstrack} shows the predicted radius evolution for a $1.2~M_\odot$ star from the \citet{Baraffe2015} models. Both figures include the star evolution between 4 and 45 Myr. 

We show the evolution of the nuclear-fusion luminosity for the p-p chain and the CNO cycle plotted as a fraction of the total stellar luminosity (Fig.~\ref{fig:cno_analysis}, top) and the evolution of the central abundance of $^3$He, $^{12}$C, and $^{14}$N {  (Fig.~\ref{fig:cno_analysis}, bottom)}.
In Fig.~\ref{fig:baraffemasstrack}, we show the evolution of the central temperature and density and the mass in the radiative core as a fraction of the highest value attained between 4 and 45 Myr. Around 13 Myr and 25 Myr, both sets of standard models predict that the stellar radius increases for a period of time as the star progresses toward the zero-age main sequence (ZAMS), reaching it at $\sim$30 Myr. 
The radius increase at around 13 Myr is associated with the ignition of p-p chain fusion through the p-p I chain
\citep[see reactions~(1)--(3) in][]{Chabrier1997}.
At this age, p-p chain fusion produces about 1\% of the total stellar luminosity. This is insufficient to stabilize the stellar core against gravitational contraction, leading to an overproduction of energy that causes the stellar envelope to expand while the core heats up and contracts. 
As long as $^3$He has not reached its equilibrium, its abundance increases with temperature, and the nuclear energy in the central region increases rapidly with time. When the third reaction of the p-p I chain 
\citep[see reaction~(3) in][]{Chabrier1997} 
becomes fast enough to efficiently destroy $^3$He, its abundance decreases after reaching a maximum (see Fig.~\ref{fig:cno_analysis} at 
$\sim$19~Myr). 
Once equilibrium is reached, the abundance of $^3$He decreases as temperature increases, as observed in Fig.~\ref{fig:cno_analysis} at $\sim$24~Myr
\citep[e.g.,][]{Clayton1968, Baraffe2018}. 
It is followed shortly thereafter by the beginning of CN cycle fusion.   
The production of nuclear energy due to the ignition of  the CN cycle yields the formation of a small convective core (as shown in Fig.~\ref{fig:baraffemasstrack}). The progression in mass of the convective core  affects the abundance of central $^3$He, resulting in the ``fusion bump'' of the $^3$He abundance observed in Fig.~\ref{fig:cno_analysis} at $\sim$21 Myr \citep[see Sect. 2.1 in][]{Baraffe2018}. 
A radius increase is again seen starting around 25 Myr. Declining energy production from the CN cycle as $^{12}$C is processed into $^{14}$N leads to a slight radius increase as the core begins to contract. Achieving thermal equilibrium is accomplished once the full CNO cycle comes into equilibrium. A critical abundance of $^{14}$N must accumulate to overcome the relatively low fusion probability between $^{14}$N and a proton.

\begin{figure}
\centering
\includegraphics[width=1.0\hsize]{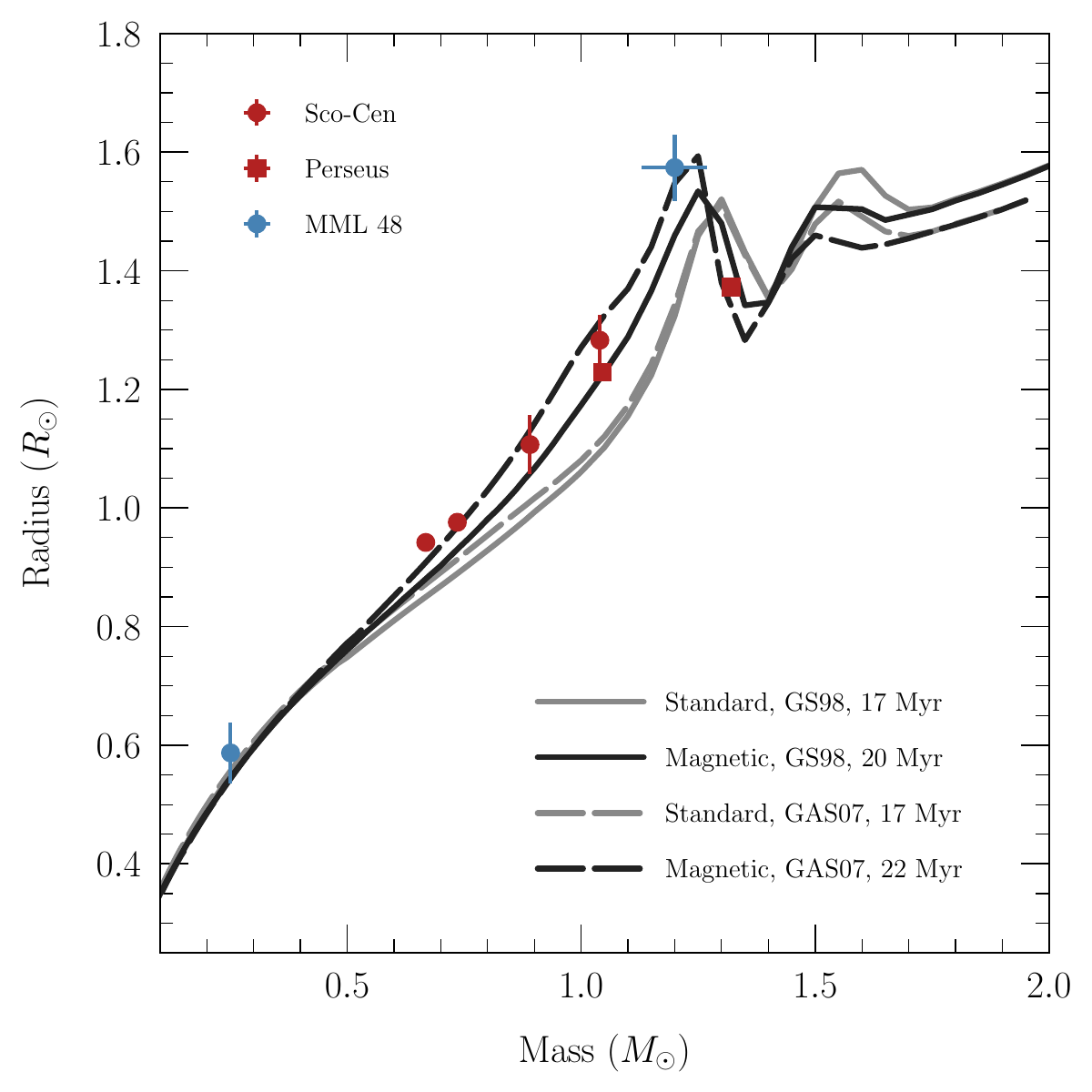}
\caption{Mass-radius diagram for low-mass stars in eclipsing binaries with intermediate ages compared to Dartmouth standard{  (gray}) and magnetic (black) models. 
The circles are for the EBs in Sco-Cen: components of 
MML~53 \citep{GomezMaqueoChew2019} and 2M1222--7 \citep{Stassun2022}  {  are shown in red},
and MML~48 (this work) {  is} given in blue. 
The red squares with error bars are for each component of NP Per \citep{Lacy2016}. The standard models show the first fusion bump at 17 Myr, but at a larger mass than the MML~48 primary. The Dartmouth magnetic models can best reproduce the ensemble of EBs, including the large radius of the MML~48 primary, but they require an age of 20--22 Myr depending on the solar composition. }
\label{fig:cnobump}
\end{figure}

Returning to Fig.~\ref{fig:mr}, a fusion bump is observed in the isochrones of \citet{Baraffe2015} that corresponds to the building of $^3$He in the core as described above. 
The enlarged radius of MML 48 A could potentially be explained by the aforementioned process if the system is slightly older. However, this presents a problem because higher mass stars in Sco-Cen suggest an age closer to 16 Myr. Alternatively, it could also be explained if the mass of{  MML 48~A} were higher, thus showing the fusion bump earlier and in closer agreement with the UCL age. 
In Fig.~\ref{fig:baraffemasstrack}, we also show that an earlier and more significant increase in radius can be attained with a higher initial Helium abundance of Y = 0.29. 
We did not attempt to explore all of the possible scenarios, because we consider model-independent stellar properties are needed beforehand.

As proof of concept, we attempted to reconcile the discrepancy using magnetic Dartmouth stellar-evolution isochrones, which have been successful in explaining intra-cluster age discrepancies \citep[e.g.,][]{Feiden2016}. Standard and magnetic Dartmouth isochrones are plotted against EBs from Sco-Cen and Perseus that have ages determined to be 16--17 Myr in Fig.~\ref{fig:cnobump}. Standard models at 17 Myr predict the first fusion bump at a higher mass than that of MML~48~A at this age. 
We find broad agreement in the mass-radius diagram with magnetic models, assuming population ages of 20 -- 22 Myr depending on the adopted solar composition. MML 48 A is predicted to occur near the peak of the first fusion bump, while NP Per A appears to be between the first and second bumps. There is still notable disagreement between models and the radius of MML~48~B, but our results provide a higher level of consistency between age estimates from the higher mass stars, including capturing relevant physics needed to explain the radius of MML~48~A.

\section{Summary} 

The EB MML~48 is an eclipsing binary system where the two eclipsing stars are still on the pre-main sequence. MML~48 is a member of the Upper Centaurus Lupus association, with position and kinematics consistent with the intermediate age population 
\citep[$\sim$16~Myr;][]{Mamajek2002,Pecaut2012,Pecaut2016,Wright2018}. The measured radii, masses, and temperatures of the eclipsing stars, presented in this paper, are based on a single-lined spectroscopic {  RV} curve making them dependent on stellar models. 

The period and time of minimum of MML~48 was challenging to measure because of systematic uncertainties attributed to stellar activity (i.e., spot modulation in the light curves). However, the very long time span of our observations and high precision of the follow-up light curves and TESS data allowed for a robust period to be derived despite these complexities. 

Fortuitously, the more massive MML~48~A star has been caught at a time where there is energy over-production in the core due to a build-up of $^3$He.{  This is the first time that a young star in an eclipsing system has been observed during
its fusion bump.}
Although we show that magnetic models can provide a better fit to the ensemble of EBs with intermediate ages  and the large difference in radius between MML~48~A and NP~Per~A (Fig.~\ref{fig:cnobump}), they cannot explain the discrepancy with the cluster ages. 
Stellar properties that are model independent are needed before more thorough and in-depth explorations of the underlying physics in the models, such as an enhanced initial helium abundance (Fig.~\ref{fig:baraffemasstrack}), are done.  In addition, fully understanding and constraining the timing of the observed fusion bump will require a main-sequence turn-on age for the cluster that is derived from the same models that are used to fit the EB properties in an internally consistent way.  Internal consistency with stellar models is important when attempting to constrain model accuracy and physics.

\begin{acknowledgements}
This work was supported by the TESS GI Cycle 3 grant ID G03143 (PI: Hebb). 
YGMC, RP and FF are partially funded by UNAM PAPIIT IG-101224. 
IB is partly supported by STFC grant ST/Y002164/1.
RP has been partially supported by the Consejo Nacional de Investigaciones Cient\'ificas y T\'ecnicas (CONICET; Argentina) through project PIBAA-CONICET ID-73811.  
MB is supported by the Ed Tapper Fund and partially supported by NSF award 1852158.
      This research has made use of the services of the ESO Science Archive Facility. 
      This work makes use of observations from the LCOGT network. 
      Based on observations at Cerro Tololo Inter-American Observatory using the 0.9m and 1.5m telescopes operated by the SMARTS Consortium out of Georgia State University.   
      The Chiron observations were supported in part by NSF Grant AST-1614113 to Stony Brook University.
Based on data acquired at Complejo Astronómico El Leoncito, operated under agreement between the Consejo Nacional de Investigaciones Científicas y Técnicas de la República Argentina and the National Universities of La Plata, Córdoba and San Juan
    WASP-South is hosted by the South African Astronomical Observatory and we are grateful for their ongoing support and assistance. Funding for WASP comes from consortium universities and from the UK's Science and Technology Facilities Council.
Nick Law for making Evryscope data of MML~48 available for us, although it was not used in this paper. 
The TESS data presented in this paper were obtained from the Mikulski Archive for Space Telescopes (MAST). STScI is operated by the Association of Universities for Research in Astronomy, Inc., under NASA contract NAS5-26555. Support for MAST for non-HST data is provided by the NASA Office of Space Science via grant NNX13AC07G and by other grants and contracts. 
This work has made use of data from the European Space Agency (ESA) mission
Gaia (\url{https://www.cosmos.esa.int/gaia}), processed by the  Gaia
Data Processing and Analysis Consortium (DPAC,
\url{https://www.cosmos.esa.int/web/gaia/dpac/consortium}). Funding for the DPAC
has been provided by national institutions, in particular the institutions
participating in the Gaia Multilateral Agreement.
    
\end{acknowledgements}


\bibliographystyle{aa}
\bibliography{references}

\end{document}